# Cepheid Abundances: Multiphase Results and Spatial Gradients

R. Earle Luck
Department of Astronomy, Case Western Reserve University
10900 Euclid Avenue, Cleveland, OH 44106-7215
rel2@case.edu

## Abstract

Parameters and abundances have been derived for 435 Cepheids based on an analysis of 1127 spectra. Results from five or more phases are available for 52 of the program stars. The latter set of stars span periods between 1.5 and 68 days. The parameters and abundances show excellent consistency across phase. For iron, the average range in the determined abundance is 0.11 from these 52 stars. For 163 stars with more than one phase available the average range is 0.07. The variation in effective temperature tracks well with phase, as does the total broadening velocity. The gravity and microturbulent velocity follow phase, but with less variation and regularity.

Abundance gradients have been derived using GAIA DR2 parallax data (GAIA Collaboration et al. 2016, 2018), as well as Bayesian distance estimates based upon GAIA DR2 from Bailer-Jones et al. (2018). The abundance gradient derived for iron is $d[Fe/H]/dR = -0.05$ dex/kpc, similar to gradients derived in previous studies.

## 1. Introduction

Cepheid variables are evolved young stars with periods from 1.5 to 70 days, masses from four to fifteen $M_\odot$, and luminosities in the range $10^3$ to $10^4$ $L_\odot$. They, along with non-variable stars with comparable parameters, are often called intermediate mass stars. These post-main-sequence stars are of spectral type F through early K, and their main-sequence progenitors are nominally B stars. Cepheid variables and their kin are of interest from a variety of standpoints. First, they are evolved objects and their surface abundances, especially carbon, nitrogen, and oxygen, reflect the changes in composition brought on by evolutionary and/or main-sequence rotational mixing processes. Second, Cepheids are an excellent laboratory to test the consistency of abundance analyses, their temperatures and gravities change, but the chemical content does not. The determination of abundances at various points in the pulsation cycle then provides vital information on the reliability of abundances. Lastly, Cepheids are bright stars which can be seen over significant distances. This property allows them to be used to map out the variation in abundance both radially and azimuthally in a significant fraction of the Milky Way disk. The Cepheid gradient specifies the endpoint of Galactic chemical evolution and thus is a fundamental quantity in all models of Galactic chemical evolution. This work will address all of these areas.

Among the first studies of carbon, nitrogen, and oxygen in intermediate-mass stars was that of Luck (1978). This study found carbon to be deficient and nitrogen enhanced as predicted by standard stellar evolution (Iben 1967). More surprisingly; oxygen, not predicted to be modified during the first-dredge-up and subsequent helium burning, was found to be subsolar: 8.70 versus the then current estimate of the solar oxygen abundance of 8.92. This situation persisted and led to the question posed by Luck & Lambert (1985): Carbon, Nitrogen, and Oxygen in Intermediate Mass Supergiants: Is Oxygen Underabundant? Analyses of CNO in supergiants have continued (Luck 2014 and references therein) with the insistent answer that local supergiants have an oxygen content of about 8.70. The answer to the question posed by Luck & Lambert in 1985 came with new solar analyses yielding a solar oxygen abundance of 8.69 (Asplund et al. 2009). However, as pointed out by Asplund et al., this result is at odds with helioseismology results which prefer the older value of 8.92. This situation is unresolved as a three-dimensional non-LTE analysis of the primary solar oxygen indicator, the [O I] line at 630.03 nm, shows within allowable uncertainties, the line can return abundances between 8.69 and 8.92 (Socas-Nararro 2015). Studies of oxygen abundances in local neighborhood giants and dwarfs (Luck 2015, 2017, 2018) support the lower solar oxygen abundance. The contribution intermediate-mass supergiants can make to this problem is the observation that it is difficult to understand a decrease in the oxygen content of stars over the past 5 Gyr as would be required if the solar oxygen abundance were 8.92.

The work of Luck & Andrievsky (2004), Kovtyukh et al. (2005), Andrievsky, Luck, & Kovtyukh (2005), and Luck et al. (2008) typify multiphase parameter and abundance studies in Cepheids. These analyses consider Cepheids with periods ranging from 2 to 68 days. The primary results are 1) the analyses return temperatures that correlate well with phase; 2) the gravities show phase related behavior as do the microturbulent velocities; and 3) the iron abundances do not show phase-related differences, the range in abundance exhibited over all phases was usually less than 0.1 dex which is less than the typical uncertainty of a single determination.

Abundance gradients in the Milky Way have been determined from a variety of objects by numerous workers — see the introduction section of Andrievsky et al. (2002a) for a synopsis of pre-2000 work. Gradients for multiple elements were found using Cepheids as the tracer starting with Andrievsky et al. (2002a), followed by Andrievsky et al. (2002b,c), Luck et al. (2003), Andrievsky et al. (2004), Luck, Kovtyukh, & Andrievsky (2006), and culminating in the study of Luck & Lambert (2011). Other recent Cepheid gradient work includes the work of Lemasle et al. (2013), Genovali et al. (2015), and da Silva et al. (2016) concentrating on the α and heavy elements. The basic gradient result is for iron. The gradient as determined from Cepheids between Galactocentric radius ($R_G$) 3 to 18 kpc is $d[Fe/H]/dR_G = -0.064 \pm 0.002$ dex kpc$^{-1}$ (Luck & Lambert 2011).

Several factors drive reconsideration of Cepheid abundances and abundance gradients at this time. The pre-2010 analyses of Luck, Andrievsky, and collaborators used plane-parallel atmospheric models drawn from Kurucz (1992) while the Luck & Lambert (2011) analysis used spherically symmetric models from Gustafsson et al. (2008). Other differences lie in the temperature calibration, Kovtyukh & Gorlova (2000) for the initial analyses versus an updated version of the Kovtyukh (2007) calibration (Kovtyukh, private communication) for Luck & Lambert (2011). Further, Luck & Lambert found a systematic offset of +0.07 in [Fe/H] between their values and those of previous abundance determinations. These differences plus other minor variations between the analyses led Luck & Lambert to not include any of the earlier work in their final gradient determinations. What is needed now is to place all of the Cepheid abundances on a common footing. In this paper, we will consider the full extent of stars and most of the spectra available to the Luck and collaborators Cepheid studies: 1127 spectra of 435 Cepheids.

Another reason behind this work is the availability of reliable parallax data for many Cepheids. Of the 435 Cepheids to be considered, parallaxes are found in GAIA DR2 (GAIA Collaboration et al. 2016, 2018) for 389 with parallaxes greater than three times the quoted uncertainty. Before DR2, only about 20 Cepheids had reliable parallaxes and distances relied upon the Period-Luminosity relation and inexact reddening and extinction estimates. As part of this work, PL derived distances will be compared to the new GAIA parallax data. Basic information about the program stars including PL distances derived from the Madore, Freedman, and Moak (2017) PL relation, as well as the GAIA DR2 parallax data, can be found in Table 1.

## 2. Observational Material

The spectroscopic data used here was used in the prior Luck and collaborators Cepheid analyses. There are four data sources:

- The Sandiford Echelle Spectrograph on the 2.1m Struve Telescope of McDonald Observatory denoted "S" in the Tables.

- The High-Resolution Spectrograph on the 2.7m Harlan J. Smith Telescope of McDonald Observatory denoted "M" in the Tables.

- The High-Resolution Spectrograph of the Hobby-Eberly Telescope denoted "H" in the Tables.

- The FEROS spectrograph on the 2.2m MPG telescope at the European Southern Observatory. These are denoted "F1" in the tables if originally from Luck et al. (2011) and "F2" if from Luck (2014).

The data reduction procedures used were described in detail in Luck (2015, 2018) for Sandiford spectra, in Luck & Lambert (2011) for the 2.7m and HET High-Resolution Spectrographs, and in Luck et al. (2011) for FEROS spectra. All spectra were processed from raw frames to 1-D spectra using IRAF. The author developed spectra reduction package ASP was used for final cosmic ray and cosmetic defect removal, B star division to remove terrestrial lines, continuum placement, wavelength setting, and equivalent width determination using the Gaussian approximation. More information about the spectra can be found in the following paragraphs. A log of all observations is given in Table 2. It provides the date, starting time, length of exposure, and phase for all observations.

The Sandiford Echelle spectrograph ((McCarthy et al. 1993) has a nominal resolution of 60000, and the setup used for Cepheid observations continuously spanned the region from about 560 to 690 nm. These spectra were used in the 2002 – 2006 era analyses, but in those analyses, terrestrial divisions were not always performed, nor were the orders coadded to maximize signal-to-noise. For this analysis, both operations were done to provide the best spectra possible. There are 603 spectra of 110 stars in this group.

The McDonald 2.7m and HET spectra are the same as those described in Luck & Lambert (2011). The spectral coverage for both is about 440 to 785 nm, but only the HET spectra are continuous. The resolution for both is 30000. There are 30 spectra of 29 stars in the McDonald 2.7m data and 316 spectra of 224 Cepheids in the HET data.

The FEROS spectra have a resolution of 48000 and continuous spectral coverage from 400 to 785 nm. There are two sets of these spectra. The first is associated with Luck et al. (2011) comprising 99 spectra of 97 stars. The second set is from Luck (2014) and has 77 spectra of 50 Cepheids.

## 3. Analysis

### 3.1 Procedures and Resources

The basic precepts used in this analysis are those used in Luck (2014, 2015). Briefly, they are as follows:

- The analysis assumes local thermodynamic equilibrium (LTE).
- Effective temperatures are determined using the line-ratio – effective temperature calibration of Kovtyukh (2007) as updated by Kovtyukh (2010, private communication). This calibration was discussed in Luck (2014). A change relative to earlier analyses is an enhanced averaging process. Each ratio in the calibration yields a temperature estimate, and the previous analyses merely averaged these values and then twice applied two-standard deviation (2-σ) clips to obtain the final temperature. The problem is outliers perturbing the data clipping by generating spuriously large standard deviations. This problem is alleviated here by fitting a Gaussian to original data and then using its standard deviation to perform the initial 2-σ clip. After this, the data was re-averaged and then 2-σ clipped twice to find the final temperature. In ratios determined from high signal-to-noise data, there is little difference between the two methods, but for lower signal-to-noise data, the difference in final temperature can be several hundred degrees. The procedure used here hopefully yields the better value.
- Gravities and microturbulent velocities are set by standard spectroscopic techniques – an ionization balance between Fe I and Fe II for gravity and by forcing there to be no dependence in the per-line Fe I abundances on equivalent width.
- Total broadening velocities are established by minimizing the chi-square of synthetic versus observed spectra using lines in the 560 to 580 nm range. All broadening profiles are Gaussian.
- The line list predecessors are the lists of Kovtyukh & Andrievsky (1999) and Luck & Heiter (2007) supplemented by clean solar lines drawn from the work of Rutten & van der Zalm (1984a,b) and other lines used in the determination of solar abundances. There are 2943 lines in the list. Equivalent widths were measured using the Delbouille, Roland, & Neven (1973) solar intensity atlas. An inverted solar analysis then provides oscillator strengths for the lines. The solar abundances assumed are those of Asplund et al. (2009) for carbon, nitrogen, and oxygen. All other elements take their abundances from Scott et al. (2015a, 2015b) and Grevesse et al. (2015). The Unsöld approximation (Unsöld 1938) is used to compute any damping constant without a more precise determination such as provided by Barklem et al. (2000) or Barklem & Aspelund-Johanson (2005). This line list was used to analyze solar reflection spectra (Luck 2018) to determine offsets in abundance from the input values. These offsets have been applied to the final abundance data presented here.
- Hyperfine structure for Mn, Co, and Cu was accounted for using the hyperfine data of Kurucz (1992).

- Syntheses of Li, C, N, and O used the atomic and molecular line data specified in Luck (2014, 2017). The pertinent features are the hyperfine Li I doublet at 670.7 nm, C I lines at 505.2, 538.0, and 710.5 nm, $C_2$ Swan system lines centered about 513.5 nm, N I lines near 745.0 nm, and the O I triplet at 615.5 nm and the [O I] line at 630.0 nm.
- Model atmospheres are interpolated from the grids of Gustafsson et al. (2008). The models used are spherically symmetric models and require a mass choice. The period-radius-mass relations of Bono et al. (2001) and Groenewegen (2007) were used to derive masses for all program Cepheids. Most of the program Cepheids have derived masses in the range 3 to 7 $M_\odot$. For model selection, the dividing line was placed at 7.5 $M_\odot$, if the mass is less than this, a 5 $M_\odot$ model is used, above this point, a 10 $M_\odot$ model. The models are all moderately CN-processed: [C/H] = −0.13 and [N/H] = +0.53. The available metallicities are [M/H] = −0.5, 0.0, and +0.25. The [M/H] = −0.5 models are α-enhanced by +0.2 dex. All models used have a turbulent velocity of 5 km s$^{-1}$.

The parameter and abundance results of the Cepheids are found in Tables 3 through 7. Table 3 contains the per phase effective temperature information – the mean temperature determined from the line-ratios, the standard deviation of the mean, and the number of line ratios used. Also given is the gravity, the microturbulent velocity, and details of the Fe I and Fe II data – mean abundance, standard deviation, and number of lines. Table 4 presents the mean [x/H] ratios for $Z > 10$ for each Cepheid on a per phase basis. For elements with both the neutral and first ionized species available, the mean is $(n_I X_I + n_{II} X_{II}) / (n_I + n_{II})$ where n refers to the number of lines, X to the mean elemental abundance, and I and II refer to the neutral and first ionized species respectively. For yttrium and zirconium in the stars with an effective temperature above 6000 K, only the ionized species are retained in the average. Table 5 has the details of the per phase $Z > 10$ abundances: mean abundance, standard deviation, and number of lines for each species considered. Table 6 has the details of the Li, CNO analysis and Table 7 enumerates the per star average abundances.

### 3.2    Parameter and Abundance Inspection

#### 3.2.1 Temperature, Gravity, Microturbulent Velocity, and [Fe/H] Inspection

Accessing the reliability of Cepheid stellar parameters and abundances on a per phase basis relies on a two-fold approach: one first inspects the internal estimates of uncertainty, and second, compares the results to those of previous analyses. For these Cepheids, the internal uncertainties for temperature and abundances are quantified in Tables 3 and 5 on a per spectrum basis. The statistics presented in Tables 3 and 5 reflect only the line-to-line variations. In Table 8, descriptive statistics for the temperatures and Fe I data are presented. For the temperatures, most standard deviations are in the range 60 to 100 K. Breaking the information down by spectrum source, tantamount to segregating by resolution and signal-to-noise, one finds the lowest uncertainties are associated with the highest signal-to-noise and best resolution data, i.e., the S, M, and F1 datasets. The H dataset has the faintest Cepheids and the lower signal-to-noise and resolution data, and thus the larger scatter about the mean values. The iron information presented in Table 8 shows the standard deviation of the mean of the Fe I lines on a per spectrum basis to be of order 0.14 dex independent of source.

Uncertainties in gravity in an ionization balance analysis are found by determining how significant a change in gravity is allowed before the total abundance of iron as determined from Fe I and Fe II are judged not to agree. In Table 9, the sensitivity of Fe I and Fe II ensembles are shown as a function of parameter changes. Assuming no change in effective temperature, and a maximum value of Fe I – Fe II of 0.05 dex, the gravity cannot change by more than 0.15 dex for the ionization balance to be maintained. The ionization balance can be maintained over substantial gravity differences if the effective temperature is allowed to vary. Lowering the temperature by about 100 K and lowering the gravity by about 0.3 dex will maintain the ionization balance – see Table 9. Additionally, such a change has only a small effect, about +0.02 dex, on the total iron abundance.

The microturbulent velocity was determined by demanding there be no dependence of the iron abundance as determined from Fe I lines on equivalent width. The uncertainty in the slope of the abundance – equivalent width relation is much smaller than what a change of 0.3 km s$^{-1}$ in the microturbulent velocity yields. This level of sensitivity implies the microturbulent velocity is formally good to 0.05 to 0.1 km s$^{-1}$. The formal value is likely an underestimate as the parameters all interact with one another. A more reasonable value is ±0.3 km s$^{-1}$.

The standard deviations of the Fe I data on a per spectrum basis average about 0.14 dex with a minimum of 0.09 dex and a maximum of 0.44 dex – see Table 8. There is no significant variation as a function of spectrum source. As will be seen in the examination of the multi-phase results, the iron abundances are more consistent than their standard deviation reflects, perhaps a better measure would be the standard error of the mean. However, this quantity seems much too low – about 0.01 dex for these stars.

External comparison of parameters and abundances are available for the majority of the spectra used here. Those analyses are from Luck and collaborators and are enumerated in Table 10 where the descriptive statistics for the comparison are presented. Comparison of the temperature, gravity, and iron abundance information shows good agreement. The mean temperature difference over the entire sample is 6 K (this work – old), with no meaningful differences shown between the various data sources. The same is true for the gravities, which show a mean difference of −0.09 dex, and the derived [Fe/H] ratios, which show a mean difference of −0.03 dex. In the microturbulent velocities, there is a difference of −0.89 km s$^{-1}$. This difference arises because the pre-2014 analyses set the microturbulent velocity by forcing the Fe II per line abundances to show no dependence on equivalent width. In Luck 2014, the F2 subset, and here, the more common method of using Fe I to set the microturbulence is used. Fe I allows utilization of a greater number of lines with a larger range in equivalent width. Luck (2014) used the Fe I lines to determine the microturbulent velocity, and those microturbulent velocities agree well with those determined here.

Iron abundances are available for a number of these Cepheids from other sources. Table 10 compares some recent analyses to the current results with the overall conclusion being that the iron data agree well. The worst agreement is with Lemasle et al. (2013) where a mean difference in [Fe/H] of +0.10 dex is found. However, the same difference is found between Lemasle et al. and Genovali et al. (2014, 2015).

### 3.2.2 Gravity versus Period-Radius-Mass Determination

Masses for all program Cepheids were derived from the period-mass-radius (PMR) relations of Bono et al. (2001) and Groenewegen (2007). The average mass for the sample is 5.5 $M_\odot$ with a standard deviation of 0.06 $M_\odot$. The total mass range found in the sample is 3.53 to 10.6 $M_\odot$. Turner (2012) derives a lower mass limit for Cepheids of 4.1 $M_\odot$ based upon the mass of SU Cas in the open cluster Alessi 95. The mass found here for SU Cas is 3.8 $M_\odot$. Given the simplicity of the technique used here, this level of agreement is adequate.

Surface gravities were derived from the masses discussed above. The gravity derived from the PMR determination is effectively an average value for the Cepheid in question. In Figure 1 (top panel), the derived per phase gravities for all of the Cepheids considered here are plotted against the PMR relations derived gravity. The individual values scatter around the line of equality indicating PMR relations give fair estimates of the mass and radius.

The comparison can be taken further using the Cepheids with multiphase data. Average gravities for 52 Cepheids are available from this dataset. The bottom panel of Figure 1 shows these values versus the PMR relations value. The mean values determined from the multiphase results are in good accord with the PMR values. A linear fit to the data yields a line mainly parallel to the line of equality. The implication is if one needs a starting value for an ionization balance, the gravity derived from the period-radius-mass relations is an excellent choice.

### 3.2.3 Li, C, N, O results

Lithium, carbon, nitrogen, and oxygen abundances have been derived using spectrum synthesis. Synthesis matching requires the total broadening velocity to be determined. The fit is done using lines in the 560 – 580 nm region. The total broadening velocity is set by forcing the lines to the observed depth at a range of broadening velocities. The best-fit velocity is determined from a $\chi$ – square minimization. These velocities are given in Table 6 along with the Li, C, N, and O abundance data.

Lithium in Cepheids is an elusive element. The hyperfine doublet at 670.7 nm is the source of all lithium information. The only star in this sample with a strong lithium presence is V1033 Cyg with an LTE lithium abundance of 3.19. Luck & Lambert (2011) first noted this star. Table 6 indicates most of these stars have only an upper limit on the lithium abundance. The procedure used to assign status is if the measured feature depth at 670.78 nm is less than 2 percent, the abundance derived from the profile is automatically considered a limit. This procedure assigns the bulk of the lithium to limit status. Next, if during an inspection of the synthesis fits the data appears more uncertain than usual, and the nominal lithium depth is greater than two percent, a status of limit is manually assigned. No non-LTE corrections were applied to the lithium data.

Four carbon indicators are available for the program stars though not all are available at all temperatures, and the varying spectral coverage means not all are available for any single spectrum. The four indicators are $C_2$ in the 513.5 nm region and the C I lines at 505.2, 538.0, and 711.3 nm. The first two C I features are single lines while the feature at 711.3 nm is comprised of multiple lines. In Figures 2 through 5, representative syntheses of these features are presented. The precision in the synthesis abundance is of order ±0.1 dex based and parameter related uncertainty is of equal magnitude. To determine the final carbon abundance, the data is weighted as follows: for T < 5000K only $C_2$ is used; for 5000K < T < 5500K, the simple average of $C_2$, C I 538.0 nm, and the best fitting abundance for C I 711.3 nm lines is used; and lastly, for T > 5500 K the average of all the C I lines is used with C I 711.3 nm given weight 2 and the other lines each weight 1. The C I feature at 505.2 nm is not used below T = 5500K as it is in the wing of a strong Fe I line and becomes inextricable from the Fe I line at this point.

Nitrogen abundances are available for Cepheids in the HET and FEROS datasets from the nitrogen doublet at 744.2, 746.8 nm. Two typical syntheses sets are shown in Figure 6. Abundances from these two high-potential permitted lines are only given for T > 5250K. The abundance precisions are of order ±0.1 from the syntheses and a similar amount from parameter uncertainties.

The oxygen abundance indicators are the O I triplet at 615.8 nm, and the [O I] line at 630.0 nm. Representative syntheses of these two features are shown in Figures 7 and 8 respectively. The [O I] line in supergiants is easily measured provided a careful terrestrial $O_2$ line cancellation be made. Unlike the Sun, where the line is less than 0.5 pm in equivalent width, the line in Cepheids is moderately strong, about 3.5 pm at 6000 K and solar abundance increasing to 5.5 pm at 5250 K. The primary sensitivity of the line is to gravity. Final oxygen abundances are weighted means of the O I and [O I] abundances: for T < 5500K only [O I] is used, for 5500 < T < 6000K the [O I] lines has weight 3 and O I weight 1, for 6000 < T < 6400K the O I lines have weight 3 and [O I] weight 1, and above 6400 K only O I is used. In the determination of the CNO abundances, the molecular equilibrium calculation includes the all relevant molecules with CNO components. These stars are sufficiently hot that the C and O abundances are not interlocked through CO.

## 4. Distances

Two ways to determine the distance to a Cepheid are now generally available, the first being the use of a PL relation coupled to a line-of-sight extinction determination, and the second a direct geometric measure of the distance, i.e., a parallax. For the PL method, a variety of information is needed. The fundamental period for the program Cepheids in Table 1 is from the GCVS (Samus et al. 2017) or specific sources as given in Table 1. The reddening used is from Turner (2016), Madore, Freedman, & Moak (2017), or from Fernie (1995) as modified by Fouqué et al. (2007). Intensity V mean values are generally taken from Fernie (1995) or as given in the references in Table 1. To convert from E(B−V) to $A_V$ the relation of Fouqué et al. is used: $R_V = A_V/E(B-V) = 3.23$ for most stars. In the Carina region, 265 < l < 315 degrees, $R_V = 4.0$ is used (Carraro et al. 2017). Turner et al. (2014) give the range of $R_V$ to be 2.8 to 4.0. An analysis of the 661.3 nm diffuse interstellar band in a subset of these Cepheids (Kashuba et al. 2016) indicates a significant variation in R as a function of galactic latitude. The variation of $R_V$ is a major uncertainty in PL derived distances. For example, at a reddening of 0.5, changing $R_V$ from 3.23 to 4.0 changes the distance modulus by 0.38 magnitudes.

In Figure 9, distances derived from the GAIA DR2 parallaxes (GAIA Colloboration et al. 2016, 2018) are shown versus the distances derived from the period-luminosity relation. What is immediately evident from the top panel of Figure 9 is while GAIA DR2 represents significant progress in the determination of parallaxes, the parallaxes for the more distant Cepheids are uncertain at a significant level. To better understand the relationship between the DR2 derived distances and the PL distances the bottom panel of Figure 9 shows those stars for which the DR2 parallax is 8 or more times the uncertainty. While the diagram is cleaner than the unfiltered distances, the apparent problem is the distances do not scatter uniformly about the line of equality. The parallax distances are generally larger than the PL distances.

A potential problem could be systematic effects in the GAIA parallax distances. Bailer-Jones et al. (2018) provide a Bayesian estimate of the true parallax distance based on the GAIA DR2 result and a weak distance prior based on the Galactic latitude and longitude. Figure 10 (top panel) shows the relation between the distance based on the parallax and the Bayesian "corrected" distance. One can see the "corrected" distances deviate systematically from the raw distances with the "corrected" distances being the smaller of the two. The variation is about 15% at five kpc and rises to about 33% at ten kpc. If one examines the behavior of the "corrected" distances versus the PL distances, one obtains the result shown in the bottom panel of Figure 10. It is obvious by comparing the top panel of Figure 9 to the bottom panel of Figure 10 that the "corrected" distances agree better with the PL distances, but significant scatter remains. A caveat about the "corrected" distances concerns those stars in GAIA DR2 with negative parallaxes. A case in point is δ Cep. In DR2, the parallax is given as −1.17 ±0.47 mas. Bailer-Jones et al. give the corrected distance as 4.35 kpc. However, the Hipparchus parallax for δ Cep is 3.77 ± 0.16 mas (van Leewen 2007) yielding a distance of 0.27 kpc. The byword here is caveat emptor. The question at this point is – what distances should be adopted to compute radial abundance gradients? The easy answer is to wait for the final GAIA parallaxes and yet better distances. What one can do at this point is to determine the sensitivity of the gradient is to differing assumptions about the distances.

Radial abundance gradients depend not only upon the distance to the individual stars but also the distance of the Sun from the Galactic center ($R_0$). In prior gradient work (see, for example, Luck & Lambert 2011), the solar distance was assumed to be 7.9 kpc (McNamara et al. 2000). Turner (2014) advocates $R_0$ in the range 8.24 to 8.34 kpc. However, Francis & Anderson (2014) determined $R_0$ to be 7.4 ±0.3 kpc based on globular cluster distances and 2MASS bulge periphery estimates. Their literature review indicates when systematic effects are taken into account, the majority of $R_0$ values determined up to that time are consistent with 7.4 kpc. However, the discussion is not over as Majess et al. (2018) give $R_0$ = 8.3 ± 0.36 kpc based on RR Lyr stars. Fortunately, the effect of $R_0$ on radial abundance gradients is a second order effect, mainly moving the Cepheid sample radially with respect to the Galactic center. Evaluating the gradients with $R_0$ varying from 7.4 to 8.3 kpc only changes the gradients by about 0.0004 dex/kpc while the gradients are of order −0.03 to −0.05 dex/kpc. All radial gradients given in this work adopt $R_0$ = 7.9 kpc.

## 5. Results and Discussion

The first topic to be taken up here will be multiphase abundances for individual Cepheids. Next, abundance trends within Cepheids will be examined, and lastly, spatial abundance trends will be considered.

## 5.1 Multiphase Results

Cepheids provide one of the most reliable tests of the reliability of parameters and abundances of any star. During their pulsation cycle, the parameters vary systematically, but the abundances cannot vary. Parameter variation and total iron abundance per phase for seven stars with periods ranging from 3.15 to 45.01 days are shown in Figures 11 through 17. The per-phase abundances shown can be found in Table 4. The seven stars illustrated are a small subset of the total multiphase data available. There are 164 stars with two or more phases available, and 52 stars have 5 or more phases available. Many Cepheids show period changes, so the use of proper light elements is critical. The phasing used here uses the most recent elements from the GCVS (Samus et al. 2017) including where available period change corrections. The stars with extensive multiphase data are primarily drawn from the Sandiford (S) data and were previously discussed by Luck & Andrievsky (2004), Kovtyukh et al. (2005), Andrievsky, Luck, & Kovtyukh (2005), and Luck et al. (2008). For more phase-dependent parameter figures, consult those papers.

Among the stellar parameters effective temperature, gravity, and microturbulent velocity, the effective temperature shows the clearest signal of the pulsation. The amplitude for these stars varies between 1000 K and 1400 K depending on the period: shorter periods have smaller amplitudes and higher overall temperatures. Two of the stars shown, SZ Tau and ζ Gem, are s-Cepheids. s-Cepheids are Cepheids with smaller light amplitudes and sinusoidal light curves (Turner 2012). The observed temperature amplitudes of 400 K for SZ Tau and 600 K for ζ Gem are consistent with this idea. The regularity of the temperature curves indicates the phasing of these Cepheids is reliable.

The gravities of the stars shown in Figures 11 through 17 (second panel down) indicate at shorter periods little evidence of a variation that correlates with phase. At periods of 15 days and greater, there is a believable gravity curve, but the variation is not as regular as is the temperature curve. One can determine the expected ratio of radius variation from the observed magnitude difference at maximum and minimum light combined with the effective temperatures at those times. For 5 to 7 day classical Cepheids, the V magnitude amplitude is about 0.8 magnitudes. For 15 day and longer period Cepheids, the amplitude is of order one magnitude in V. Ignoring the bolometric correction; the luminosity variation is in the range of a factor of 2 to 2.5. The 5 to 7 day Cepheids vary from about 5000 to 6000 K which leads to an energy output variation of about a factor of two. Similarly, the temperature variation in longer period Cepheids gives rise to luminosity variation of about a factor of 2.5. Thus, one would not expect a significant gravity variation because of radius changes. The gravity changes result from a coupling of the dynamical term proportional to dV/dt to the radius change.

Microturbulent velocities most often show a correlation with phase. The maximum velocity appears to occur somewhat before maximum temperature by about 0.1 to 0.3 in phase. While the variations are not large and are comparable to the precision estimates, the trends are consistent within each dataset, and from Cepheid to Cepheid. This behavior was perhaps first noted by Mel'nikov (1950) and was studied extensively by van Paradijs (1971) who showed the maximum microturbulent velocity is attained at about minimum radius, i.e., at maximum temperature.

The total broadening velocities are shown in the next to bottom panel of Figures 11 – 17. The broadening velocity as defined here is the Gaussian macroturbulent velocity needed to match the line profiles. There are two items of note here; the first is the velocities are rather large; the smallest noted is about 5 km s$^{-1}$ and the largest 24 km s$^{-1}$. Next, the velocities correlate with phase with the maximum velocity occurring at, or just before, maximum temperature. There does appear to be a correspondence between the phase behavior of the microturbulent velocity and the total broadening velocity.

Turbulence in Cepheid atmospheres has been examined by numerous studies – see for example Benz & Mayor (1982), Breitfellner & Gillet (1993a,b,c), or Bersier & Burki (1996). The time behavior of microturbulence in δ Cep was examined by Gillet et al. (1999) who demonstrate that global compression and rarification of the atmosphere drives the changes in microturbulence. Additionally, rapid changes in turbulence are induced by the shock waves associated with the compression or rarification. However, as stated by Gillet (2014), while the essential features of Cepheid pulsation are well-understood, the detailed dynamics of a Cepheid atmosphere are still unknown.

The bottom panel of Figures 11 through 17 shows the iron abundance as a function of phase. The hope is iron will show no dependence on phase, and overall, this expectation is borne out by the behavior of the data. What is interesting is the spread evidenced in the iron abundances. A typical standard deviation of a per-phase Fe I ensemble is 0.14 dex – this is the error bar shown with the data. The iron data exhibited in the figures typically shows a total range of 0.07 dex, just over half of the standard deviation of an individual determination. This agreement over phase attests to the internal consistency of the analysis.

Multiphase data allows the determination of average temperatures, gravities, microturbulent velocities, and total broadening velocities. Fifty-two Cepheids in the current study have sufficient data for the computation of these parameters – see Table 11. The average quantities are shown versus log (Period) in Figure 18. The behavior of the temperature, gravity, microturbulence, and total broadening velocity is systematic with respect to log (P). The mean temperature decreases with increasing period, gravity decreases, and the two velocities both increase. In the Cepheid total broadening velocity data, a subset having higher mean velocities than most Cepheids of similar period is apparent. Nothing stands out about these stars other than their more substantial broadening.

A point to be made about the total broadening velocity is that its systematic behavior with respect to period points to an origin in macroturbulence, not rotation. One would not expect the rotation to correlate with period given the random inclination of these objects to the line-of-sight. Additionally, at the velocities observed, rotation and macroturbulent profiles are separable, and there is no indication of the need for the inclusion of rotation to fit the line profiles in any of these stars.

In the bottom panel of Figure 18, the average iron abundances are shown versus log (Period). This data shows the iron abundance does not depend on period. The error bars in this panel are not one standard deviation about the mean; they are the total range in iron content – these iron abundances are very well determined.

## 5.2     Abundances Trends

Before proceeding to a discussion of spatial abundance gradients, a discussion of several relevant aspects of the abundances is warranted. An exploratory tool often used in the discussion of abundances, and in particular, the [x/Fe] ratios, is the distribution of those ratios with respect to atomic number. In Figure 19, this information is displayed. For the $Z > 10$ elements, the [x/Fe] ratios have been corrected using solar abundances from Luck (2018) derived from reflection spectra. Solar reflection spectra were shown by Luck to yield the canonical solar values for carbon, nitrogen, and oxygen (Asplund et al. 2009) using the atomic and molecular data used here. The salient features of the <[x/Fe]> ratios are a carbon deficiency linked with a nitrogen enhancement, a sodium enhancement, a general overabundance level of 0.1 dex in the α- and iron-peak elements, and a mild overabundance of the rare earths and lanthanides. α- and heavier elements overabundances are reminiscent of dwarfs in the local neighborhood – see Luck (2018). In the local dwarfs, the overabundance level relative to the Sun is about 0.05 to 0.1 dex – somewhat less than seen in the Cepheids. The overabundance level in the Cepheids could point in one of two directions, a systematic error affecting the abundances, or the preferred alternative, the progression of Galactic chemical evolution over the past several gigayears.

The carbon and nitrogen behavior is easily understood as the result of the first dredge-up (Iben 1967). During the first dredge-up, the results of incomplete CN-processing are mixed to the surface. The upwelling material is carbon-poor, nitrogen-rich leading to the observed surface composition. The sum of C+N is preserved in the process. In these stars, the mean <[(C+N)/Fe]> is 0.01 dex in accord with the predictions of standard stellar evolution. However, while the picture presented above is comforting, it is not the entire story.

The main-sequence progenitors of Cepheids are B0 to B5 stars. Early B stars show a large range in projected rotational velocity, and among the B stars with lower velocities – less than 100 km s$^{-1}$, the nearby examples show essentially solar CNO abundances (Lyubimkov et al. 2013 and references therein). However, in rapid rotating early B stars there exists the possibility for surface CN composition changes resulting from rotationally induced mixing. If this happens, the surface CN composition of post-first giant branch stars, explicitly including Cepheids, reflect the combination of the rotationally mixed material with the material of the first dredge-up. This option was examined by Lyubimkov et al. (2011) using the models of Heger & Langer (2000) and Maeder et al. (2009) to investigate the nitrogen abundances of non-variable supergiants of spectral type A and F. Their conclusion is the better fit to the observed nitrogen abundances comes from the evolution of lower rotational velocity B stars through the first giant branch. They point out the dominant contribution to the CN rearrangement comes from the first-dredge-up. They acknowledge the observed nitrogen abundances are compatible with changes predicted from rotational mixing in high rotation stars making the first blue-to-red crossing of the instability strip. However, given that all of the A and F supergiants exhibit similar nitrogen abundances, a timescale argument eliminates the second possibility. The argument is the time spent in the helium core-burning phase (post first giant branch) is much longer than the time to cross the instability strip in the first blue-to-red crossing. Since the nitrogen abundances found here are consistent with those found by Lyubimkov et al., Cepheid compositions will result from the same process.

In Figure 20 (top panel), the ratio [O/Fe] is shown versus [Fe/H]. The behavior exhibited here is an increase of [O/Fe] with decreasing [Fe/H]. In discussions of dwarf or giant abundances, the explanation would be this represents Galactic chemical evolution – the older stars, i.e., those with lower [Fe/H] have higher [O/Fe] ratios attributable to the increased importance of Type II supernovae oxygen production at previous epochs. However, this is not the case here as these are all young evolved stars with masses more than about four solar masses. Given that a spatial metallicity gradient does exist in these Cepheids, the less metal-rich Cepheids are preferentially located in the outer reaches of the Milky Way. Their spatial location could mean they display a somewhat different integrated chemical history relative to Cepheids inward of the solar circle, and thus show higher [O/Fe] ratios. However, the behavior of [C/Fe] as a function of [Fe/H] (middle panel of Figure 20) is not consistent with this view. As carbon and oxygen synthesis takes place in the same environment, one could expect them to show similar behavior, and they do not. Perhaps, if at lower metallicities carbon could be more depleted than expected based on standard mixing, one could rectify the behavior. One would then expect to see larger [N/Fe] ratios at lower [Fe/H], but this is not the case – see the bottom panel of Figure 20.

The sodium overabundance noted in Figure 19 is a feature of luminous stars (Sasselov (1986), Luck (1994), Luck & Wepfer (1995)) and Sasselov interpreted it as the result of the operation of the NaNe cycle. The manner in which the NaNe cycle ostensibly manifests itself in abundance is through the dependence of O/Na on mass. The proxy most readily available for mass in this sample of Cepheids is period. In Figure 21, the ratio [O/Na] is shown versus log (Period). No relationship between the [O/Na] ratio and the period is evident. Another possible reason for the behavior of sodium is that the abundance is perturbed by NLTE effects. However, NLTE calculations (Boyarchuk et al. 1988) indicate as long as the resonance doublet is avoided, the abundances should be secure. The remaining, and conceivably better possibility is the sodium content is modified during the main-sequence phase through a deep mixing process such as proposed by Denisssenkov (1994, 2005). However, Smiljanic et al. (2018) question the accuracy of current theoretical models of main-sequence deep-mixing models saying the predicted surface composition changes far exceed the observed abundances. This conclusion affects not only the NaNe deep-mixing results but also the rotational mixing events affecting carbon and nitrogen.

### 5.3   Spatial Abundance Gradients and the Local Current Metallicity

The radial abundance gradient is a critical element in galactic chemical evolution. Cepheids and other young objects set the current metallicity level and the gradient all models of chemical evolution strive to match – see for example Huang et al. (2015) or Schönrich & McMillian (2017). Previous studies have determined the radial gradient in the Milky Way to be d[Fe/H]/dR$_G$ = −0.06 dex kpc$^{-1}$ (Luck & Lambert 2011). Gradients for all other elements were found to be of comparable magnitude. In Figures 22 and 23, the iron gradient is shown for four different distance considerations. In Figure 22(a), the distances are those Cepheids from GAIA DR2 having π > 0 while 22(b) shows the same Cepheids except the distances are Bayesian estimates from Bailer-Jones et al. (2018) based on the GAIA DR2 parallax values. Figure 23(a) shows the subset of the Cepheids with GAIA DR2 parallax based distances limited to stars with parallaxes larger than five times the quoted uncertainty. Lastly, Figure 23(b) shows the Cepheid sample with PL based distances. The PL is from Madore, Freedman, & Moak (2017). In all linear fits, the data is limited to the distance range shown in plotted fit, and several stars have been excluded. The excluded stars are HK Cas, BC Aql, QQ Per, EK Del (below the abundance scale shown), and FQ Lac. EK Del, QQ Per, and BC Aql are Type II Cepheids (aka W Vir stars) and thus not relevant to the Type I Cepheid gradient. The gradients range from −0.0395 to −0.0536 dex kpc$^{-1}$, somewhat smaller than found by Luck & Lambert, but consistent with previous work. For the most consistent comparison with Luck & Lambert, the PL result is preferred. For this choice, the comparison is −0.0539 versus −0.062. The uncertainty in both values is at the 0.002 level. The difference stems from the additional stars considered here; this work uses 411 stars while Luck & Lambert utilized 313. Other gradient estimates agree well with Luck & Lambert (2011) and hence this work – see for example Genovali et al. (2014) or Anders et al. (2017). An additional "fit" is also given for each of the gradients shown in Figures 22 and 23. The fit is a lowess smoothing of the data. Over the better part of the distance range considered, the lowess fit is consistent with the linear fit. However, inward of the solar circle, the lowess smoothing indicates an upturn in the iron abundances. While the formal errors on the various iron gradients indicate they are statistically different, the reality is there is little difference between them. The favored value is from the Bailer-Jones et al. distances – d[Fe/H]/dR$_G$ = −0.0508 dex kpc$^{-1}$.

Abundance gradients for all available species are found in Table 12 with gradients for both [x/H] and [x/Fe] given. These gradients use the Bayesian distance estimates (Bailer-Jones et al. 2018) based on the GAIA DR2 parallax data. This choice is made assuming the Bayesian corrections provide a more realistic distance than the raw GAIA DR2 parallaxes or the PL distances. Stars not included in the gradient determinations are those with negative DR2 parallaxes, and those stars mentioned in the preceding paragraph. In Figure 24, the gradients are shown as a function of element. What is immediately apparent is the gradients up through the Fe-peak are all consistent with a value of about d[x/H]/dR$_G$ ≈ −0.05 dex/kpc. The gradients in the heavier elements are generally shallower with typical values of about −0.02 dex/kpc. For the gradients in [x/Fe], elements up to about Zn have no gradient in d[x/Fe]/dR$_G$ while the heavier elements show an increase in abundance relative to iron at larger Galactocentric radii. This behavior was also found in the analysis of daSilva et al. (2016).

As the last point, the current metallicity level of the local region is considered. For this exploration, the [Fe/H] ratios of Cepheids within one kpc of the Sun are considered. This sample numbers 47 Cepheids with a mean [Fe/H] ratio of +0.04 dex. The standard deviation of the sample is 0.10 dex. If one considers Cepheids in the Galactocentric radius range 7.4 to 8.4 kpc, this is, ±0.5 kpc from the solar radius, one obtains a sample of 107 Cepheids with <[Fe/H]> = +0.05 dex with a standard deviation of 0.10. The one standard deviation uncertainty in an individual [Fe/H] determination is about 0.14 dex implying the actual dispersion in the local region Cepheid abundances is very low. For comparison, the mean [Fe/H] value for local dwarfs as computed from the analysis of Luck (2018) is −0.11 dex with a standard deviation of 0.33. However, this is not a fair comparison as the dwarf sample is heterogeneous in age and origin. Local giants might provide a better comparison, but the analysis of Luck (2015) shows the giants are also inhomogeneous in age and composition. For completeness, the mean [Fe/H] ratio of the Luck giant sample is −0.05 with a standard deviation of 0.25. If one is looking for an indicator of intrinsic dispersion in a stellar generation, it appears Cepheids are the more reliable choice.

## Concluding Remarks

What does the future hold for Cepheid studies regarding stellar physics, abundance determinations, and Galactic chemical evolution? For stellar physics, the next step is extending the hydrodynamic studies of Vasilyev et al. (2017, 2018) over a broader range of parameters. Also needed are extended NLTE studies. However, these efforts will be computationally expensive and not applicable at the current time to large-scale surveys such as this one. The user-friendliness of the NLTE codes and the lack of availability to the community of input data, specifically model atoms, needs to be addressed.

A problem in abundance determinations addressable immediately is: Is there a problem with HK Cas and FQ Lac? Both appear to be metal-rich, but lie at large Galactocentric distances and thus deviate from the general trend in abundance – see Figure 23. If the abundances found here are confirmed, the implication is that the Milky Way has an inhomogeneous outer disk.

The outer region of the Milky Way disk needs a systematic exploration of abundances from high-resolution analyses. In particular, more stars in the 12 to 20 kpc Galactocentric radius range are needed. However, this is not possible with Cepheid variables – most of them are already known at this distance in the direction of the Galactic anti-center. Perhaps the answer is to explore the region with less luminous stars; i.e., giants, with the upcoming generation of large telescopes.

Acknowledgments

Financial support by Case Western Reserve University made possible the McDonald Observatory observations used in this work. I thank the Resident Astronomers of the Hobby-Eberly Telescope (HET) for their able help in obtaining these observations. The HET is a joint project of the University of Texas at Austin, the Pennsylvania State University, Stanford University, Ludwig-Maximilians-Universität München, and Georg-August-Universität Göttingen. The HET is named in honor of its principal benefactors, William P. Hobby and Robert E. Eberly. The Sandiford and HET echelle spectra used here are available through the [FGK Spectral Library](#).

This research has made use of the SIMBAD database, operated at CDS, Strasbourg, France.

This work has made use of data from the European Space Agency (ESA) mission Gaia (https://www.cosmos.esa.int/gaia), processed by the Gaia Data Processing and Analysis Consortium (DPAC, https://www.cosmos.esa.int/web/gaia/dpac/consortium). Funding for the DPAC has been provided by national institutions, in particular, the institutions participating in the Gaia Multilateral Agreement.

Table 1

Program Stars

| Name | l | b | Type | P | Mode | <V> | E(B-V) | MFM E(B-V) | T E(B-V) | d - PL | π | e_π | π/e_π | d | dmin | dMax | $R_G$ - PL | $R_G$ - π | $R_G$ - BJ |
|---|---|---|---|---|---|---|---|---|---|---|---|---|---|---|---|---|---|---|---|
| | deg | deg | | (days) | | (mag) | (mag) | (mag) | (mag) | (pc) | (mas) | (mas) | | (pc) | (pc) | (pc) | (kpc) | (kpc) | (kpc) |
| X Sgr | 1.1663 | 0.2093 | DCEP | 7.0127 | F | 4.55 | 0.237 | 0.281 | 0.252 | 284 | 3.4314 | 0.2020 | 16.9902 | 291 | 274 | 310 | 7.628 | 7.609 | 7.609 |
| W Sgr | 1.5758 | -3.9796 | DCEP | 7.5950 | F | 4.668 | 0.111 | 0.177 | 0.133 | 374 | 1.1798 | 0.4125 | 2.8598 | 1079 | 604 | 4523 | 7.551 | 7.055 | 6.824 |
| AV Sgr | 7.5336 | -0.5936 | DCEP | 15.4113 | F | 11.39 | 1.206 | … | 1.125 | 2758 | 0.5510 | 0.0689 | 7.994 | 1748 | 1550 | 2002 | 5.487 | 6.106 | 6.171 |
| AP Sgr | 8.1142 | -2.4378 | DCEP | 5.0579 | F | 6.96 | 0.178 | … | 0.217 | 761 | 1.1190 | 0.0527 | 21.2402 | 874 | 835 | 916 | 7.103 | 7.017 | 7.037 |
| VY Sgr | 10.1258 | -1.0846 | DCEP | 13.5581 | F | 11.51 | 1.221 | … | 1.192 | 2463 | 0.3895 | 0.0739 | 5.2722 | 2424 | 2027 | 3004 | 5.594 | 5.392 | 5.530 |
| ⋮ | ⋮ | ⋮ | ⋮ | ⋮ | ⋮ | ⋮ | ⋮ | ⋮ | ⋮ | ⋮ | ⋮ | ⋮ | ⋮ | ⋮ | ⋮ | ⋮ | ⋮ | ⋮ | ⋮ |

| Column | Unit | Description |
|---|---|---|
| l | degrees | Galactic longitude from SIMBAD |
| b | degrees | Galactic latitude from SIMBAD |
| Type | | Variable type |
| P | days | Fundamental period |
| Mode | | Pulsation Mode: F = Fundamental, 1 = 1st Overtone, 2 = 2nd overtone |
| <V> | mag | Mean Visual magnitude |
| E(B-V) | mag | Color excess in B-V |
| MFM E(B-V) | mag | Color excess in B-V from Madore, Freedman, & Moak (2017) |
| T E(B-V) | mag | Color excess in B-V from Turner (2016) |
| d – PL | pc | Distance using Madore, Freedman, & Moak PL. P, $A_V$ = 3.23 E(B-V) or 4.0 * E(B-V) for Carina - 260 < l < 315 degrees (Carraro et al. 2017), and $R_0$ = 7.9 kpc. E(B-V) preference order is Turner (2016), MFM (2017), E(B-V) from column 8. |
| π | mas | Parallax from GAIA DR2 |
| e_π | mas | Uncertainty in parallax |
| π/e_π | | Ratio of parallax to uncertainty in parallax |
| d | pc | Distance from Bailer-Jones et al (2018) |
| dmin | pc | Minimum distance from Bailer-Jones et al (2018) |
| dmin | pc | Maximum distance from Bailer-Jones et al (2018) |
| RG - PL | kpc | Galactocentric distance using the PL distance. |
| RG - π | kpc | Galactocentric distance computed using the GAIA DR2 parallax. |
| RG - BJ | kpc | Galactocentric distance computed using the Bailer- Jones et al. (2018) distance. |

Sources:

| | |
|---|---|
| Type | From the General Catalog of Variable Stars (Samus et al (2017)).  Individual Stars - Wils & Greaves (2004) for GSC 3725-0174 and V1397 Cyg; Berdnikov (2008) for V1359 Aql; Antipin (1997) for V458 Sct; Szabados (2006) for V411Lac, Klagyivik & Szabados (2009) for V335 Pup, LR TrA, and V340 Nor. |
| P, Mode | From the General Catalog of Variable Stars (Samus et al (2017)).  Individual Stars - Wils & Greaves (2004) for GSC 3725-0174 and V1397 Cyg; Berdnikov (2008) for V1359 Aql, MU Cep, and V382 Car; Wallerstein, Kovtyukh, & Andrievsky (2008) for QQ Per |
| <V>, E(B-V) | Primary source is Fernie (1995).  E(B-V) modified by Fouqué et al. (2007).  These values are quoted from Andrievsky et al. 2002a,b,c; Luck et al. 2003; Andrievsky et al. 2004; Luck & Andrievsky 2004; Kovtyukh et al. 2005; Andrievsky et al. 2005; Luck et al. 2006; Luck et al. 2008; Luck & Lambert 2011; Luck et al 2011; Luck 2014 |

Table 2
Observing Log

| Star | Source | Key | Date | UT Time | Exp (s) | <S/N> | Phase |
|---|---|---|---|---|---|---|---|
| AN Aur | S | anaur | 2004-10-29 | 8:52:33.37 | 3600 | 81 | 0.137 |
| AP Sgr | S | apsgr | 2004-08-05 | 4:08:39.81 | 900 | 238 | 0.125 |
| AS Per | S | asper | 2003-10-17 | 10:29:01.57 | 2700 | 130 | 0.592 |
| AW Per | S | awpera | 1996-10-23 | 9:32:40.28 | 1800 | 242 | 0.779 |
| | S | awperb | 1996-10-24 | 10:01:35.16 | 1800 | 288 | 0.937 |
| | S | awperc | 1996-10-25 | 10:49:57.40 | 1800 | 188 | 0.097 |
| | S | awperd | 1996-10-26 | 8:44:53.78 | 1500 | 224 | 0.238 |
| | S | awpere | 1996-10-27 | 8:58:14.16 | 1500 | 297 | 0.394 |
| | S | awperg | 1997-02-03 | 3:24:40.12 | 1200 | 222 | 0.674 |
| | S | awperh | 1997-10-14 | 9:46:49.92 | 1200 | 233 | 0.858 |
| | S | awperi | 1997-10-15 | 8:49:30.39 | 1200 | 358 | 0.006 |
| | S | awperj | 1997-10-19 | 9:27:50.99 | 1200 | 279 | 0.629 |
| | S | awperk | 1998-08-31 | 11:17:51.49 | 900 | 139 | 0.530 |
| ⋮ | ⋮ | ⋮ | ⋮ | ⋮ | ⋮ | ⋮ | ⋮ |

| Source | Spectrograph | References |
|---|---|---|
| S | McDonald Observatory 2.1m and Sandiford Echelle | Andrievsky et al. 2002a,b,c |
| | | Luck et al. 2003 |
| | | Andrievsky et al. 2004 |
| | | Luck & Andrievsky 2004 |
| | | Kovtyukh et al. 2005 |
| | | Andrievsky et al. 2005 |
| | | Luck et al. 2006 |
| | | Luck et al. 2008 |
| H | HET - HRS | Luck & Lambert 2011 |
| M | McDonald 2.7m | Luck & Lambert 2011 |
| F1 | ESO/FEROS | Luck et al. 2011 |
| F2 | ESO/FEROS | Luck 2014 |
| Key | Identifier | |

Note: Phase computed using GCVS elements (Samus et al. 2017)

Table 3
Cepheid Per Phase Parameters

| Star | Source | Key | Phase | T | σ | N | G | Vt | Fe I | σ | N | Fe II | σ | N |
|---|---|---|---|---|---|---|---|---|---|---|---|---|---|---|
| | | | | K | | | cm/s^2 | km/s | log (ε) | | | log (ε) | | |
| AN Aur | S | anaur | 0.137 | 5689 | 142 | 51 | 1.53 | 2.99 | 7.27 | 0.13 | 183 | 7.27 | 0.10 | 12 |
| AP Sgr | S | apsgr | 0.125 | 6310 | 35 | 45 | 1.98 | 2.81 | 7.60 | 0.13 | 181 | 7.60 | 0.12 | 17 |
| AS Per | S | asper | 0.592 | 5558 | 41 | 50 | 1.64 | 2.88 | 7.51 | 0.12 | 207 | 7.51 | 0.15 | 17 |
| AW Per | S | awpera | 0.779 | 5663 | 55 | 49 | 1.90 | 3.62 | 7.43 | 0.15 | 173 | 7.44 | 0.20 | 14 |
| | S | awperb | 0.937 | 6466 | 39 | 43 | 2.24 | 3.03 | 7.55 | 0.16 | 153 | 7.55 | 0.14 | 13 |
| | S | awperc | 0.097 | 6380 | 78 | 46 | 1.92 | 2.88 | 7.56 | 0.16 | 154 | 7.56 | 0.12 | 13 |
| | S | awperd | 0.238 | 6073 | 50 | 49 | 1.95 | 2.83 | 7.56 | 0.12 | 187 | 7.56 | 0.09 | 12 |
| | S | awpere | 0.394 | 5873 | 52 | 51 | 1.82 | 2.63 | 7.56 | 0.11 | 188 | 7.56 | 0.09 | 13 |
| | S | awperg | 0.674 | 5520 | 35 | 55 | 1.80 | 3.04 | 7.49 | 0.13 | 191 | 7.49 | 0.11 | 15 |
| | S | awperh | 0.858 | 6041 | 282 | 51 | 2.00 | 3.71 | 7.58 | 0.14 | 138 | 7.58 | 0.10 | 13 |
| | S | awperi | 0.006 | 6597 | 69 | 42 | 1.86 | 2.96 | 7.51 | 0.13 | 145 | 7.51 | 0.11 | 15 |
| | S | awperj | 0.629 | 5535 | 35 | 53 | 1.74 | 2.91 | 7.50 | 0.12 | 182 | 7.50 | 0.10 | 12 |
| | S | awperk | 0.530 | 5603 | 47 | 56 | 1.57 | 2.82 | 7.53 | 0.13 | 182 | 7.53 | 0.12 | 15 |
| ⋮ | ⋮ | ⋮ | ⋮ | ⋮ | ⋮ | ⋮ | ⋮ | ⋮ | ⋮ | ⋮ | ⋮ | ⋮ | ⋮ | ⋮ |

| Column | Unit | Description |
|---|---|---|
| Star | | Name of Cepheid |
| Source | | Source of Data / Original Paper - See source key in Table 2 |
| Key | | Identifier |
| Phase | | Phase of the Cepheid for the parameter determination |
| T | K | Effective Temperature |
| σ | | Standard deviation of effective temperature about the mean |
| N | | Number of line ratios (Kovtyukh 2007) used in the temperature determination |
| G | cm/s^2 | Logarithm of the surface gravity |
| Vt | km/s | Mictroturbulent velocity |
| Fe I | log (ε) | Abundance of iron as given by Fe I. Relative to log (ε) of H = 12 |
| σ | | Standard deviation about the mean Fe I abundance. |
| N | | Number of lines used to determine the Fe I abundance |
| Fe II | log (ε) | Abundance of iron as given by Fe II. Relative to log (ε) of H = 12 |
| σ | | Standard deviation about the mean Fe II abundance. |
| N | | Number of lines used to determine the Fe II abundance |

Table 4
[x/H] for Z>10

| Star | Source | Phase | T | G | V | Na | Mg | Al | Si | S | Ca | Sc | Ti | V | Cr | Mn | Fe | Co | Ni | Cu | Zn | Y | Zr | Ba | La | Ce | Nd | Sm | Eu |
|---|---|---|---|---|---|---|---|---|---|---|---|---|---|---|---|---|---|---|---|---|---|---|---|---|---|---|---|---|---|
| AN Aur | S | anaur | 0.137 | 5689 | 1.53 | 2.99 | 0.12 | -0.05 | 0.01 | -0.04 | -0.14 | -0.03 | 0.18 | -0.04 | -0.03 | 0.00 | -0.47 | -0.20 | -0.15 | -0.26 | -0.19 | -0.19 | 0.08 | 0.58 | 0.16 | -0.05 | 0.07 | -0.22 | -0.05 |
| AP Sgr | S | apsgr | 0.125 | 6310 | 1.98 | 2.81 | 0.73 | 0.19 | 0.33 | 0.30 | 0.32 | 0.31 | 0.62 | 0.21 | 0.16 | 0.32 | -0.06 | 0.13 | 0.38 | 0.06 | 0.54 | 0.06 | 0.34 | | 0.23 | 0.08 | 0.14 | | 0.17 |
| AS Per | S | asper | 0.592 | 5558 | 1.64 | 2.88 | 0.54 | 0.03 | 0.41 | 0.20 | 0.13 | 0.13 | 0.32 | 0.09 | 0.03 | 0.21 | -0.16 | 0.04 | 0.04 | -0.03 | -0.03 | 0.10 | 0.29 | 0.46 | 0.24 | 0.01 | -0.08 | -0.19 | -0.01 |
| AW Per | S | awpera | 0.779 | 5663 | 1.90 | 3.62 | 0.62 | 0.43 | 0.30 | 0.08 | -0.03 | 0.06 | 0.21 | 0.03 | 0.11 | 0.25 | -0.30 | -0.04 | 0.03 | 0.01 | | 0.42 | 0.36 | 0.30 | 0.19 | -0.04 | 0.21 | 0.05 | 0.11 |
| | S | awperb | 0.937 | 6466 | 2.24 | 3.03 | 0.35 | 0.44 | 0.12 | 0.23 | 0.29 | 0.29 | 0.51 | 0.15 | 0.31 | 0.38 | 0.33 | 0.08 | 0.50 | 0.00 | | 0.06 | 0.24 | | 0.33 | 0.09 | 0.43 | -0.28 | 0.13 |
| | S | awperc | 0.097 | 6380 | 1.92 | 2.88 | 0.33 | 0.54 | 0.16 | 0.25 | 0.31 | 0.34 | 0.64 | 0.17 | 0.33 | 0.34 | 0.12 | 0.09 | 0.41 | 0.03 | | 0.08 | 0.16 | | 0.31 | 0.18 | 0.30 | | 0.02 |
| | S | awperd | 0.238 | 6073 | 1.95 | 2.83 | 0.40 | 0.35 | 0.25 | 0.22 | 0.23 | 0.25 | 0.58 | 0.15 | 0.22 | 0.32 | -0.04 | 0.09 | 0.31 | 0.07 | | 0.12 | 0.48 | | 0.23 | 0.09 | 0.20 | | 0.22 |
| | S | awpere | 0.394 | 5873 | 1.82 | 2.63 | 0.54 | 0.47 | 0.32 | 0.25 | 0.27 | 0.20 | 0.54 | 0.17 | 0.17 | 0.34 | 0.01 | 0.09 | 0.13 | 0.08 | | 0.20 | 0.61 | 0.75 | 0.35 | 0.12 | 0.19 | | 0.16 |
| | S | awperg | 0.674 | 5520 | 1.80 | 3.04 | 0.63 | 0.38 | 0.35 | 0.23 | 0.26 | 0.21 | 0.21 | 0.09 | 0.02 | 0.16 | -0.13 | 0.02 | 0.15 | -0.02 | -0.17 | 0.27 | 0.34 | 0.22 | 0.43 | 0.18 | 0.22 | -0.11 | 0.24 |
| | S | awperh | 0.858 | 6041 | 2.00 | 3.71 | 0.65 | 0.13 | 0.37 | 0.29 | 0.48 | 0.11 | 0.63 | 0.29 | 0.14 | 0.41 | 0.17 | 0.11 | 0.14 | 0.06 | 0.33 | 0.41 | 0.44 | | 0.51 | 0.10 | 0.69 | 0.36 | 0.30 |
| | S | awperi | 0.006 | 6597 | 1.86 | 2.96 | 0.56 | -0.08 | 0.02 | 0.32 | 0.30 | 0.21 | 0.61 | 0.22 | 0.43 | 0.60 | 0.49 | 0.04 | 0.35 | 0.18 | 0.74 | -0.01 | 0.19 | | 0.48 | 0.31 | -0.17 | 0.26 | | 0.03 |
| | S | awperj | 0.629 | 5535 | 1.74 | 2.91 | 0.49 | -0.10 | 0.29 | 0.21 | 0.19 | 0.10 | 0.47 | 0.10 | 0.03 | 0.21 | -0.17 | 0.03 | 0.02 | -0.03 | -0.11 | 0.42 | 0.36 | 0.19 | 0.47 | 0.14 | 0.17 | -0.07 | 0.11 |
| | S | awperk | 0.530 | 5603 | 1.57 | 2.82 | 0.49 | 0.18 | 0.30 | 0.25 | 0.22 | 0.16 | 0.32 | 0.14 | 0.07 | 0.22 | -0.16 | 0.06 | 0.11 | 0.08 | 0.00 | 1.10 | 0.42 | 0.36 | 0.44 | 0.09 | 0.10 | 0.02 | 0.12 |
| ⋮ | ⋮ | ⋮ | ⋮ | ⋮ | ⋮ | ⋮ | ⋮ | ⋮ | ⋮ | ⋮ | ⋮ | ⋮ | ⋮ | ⋮ | ⋮ | ⋮ | ⋮ | ⋮ | ⋮ | ⋮ | ⋮ | ⋮ | ⋮ | ⋮ | ⋮ | ⋮ | ⋮ | ⋮ | ⋮ |

| Column | Unit | Description |
|---|---|---|
| Star | | Primary ID as given by SIMBAD |
| Source | | Source of Data / Original Paper - See source key in Table 2 |
| Key | | Identifier |
| Phase | | Phase of the observation |
| T | K | Effective Temperature |
| G | cm s-2 | log of the surface acceleration due to gravity |
| Vt | km s-1 | Microturbulent velocity |
| Na | Solar | The abundance of sodium given logarithmically with respect to the solar value. |
| Mg | Solar | The abundance of magnesium given logarithmically with respect to the solar value. |
| Al | Solar | The abundance of aluminum given logarithmically with respect to the solar value. |
| Si | Solar | The abundance of silicon given logarithmically with respect to the solar value. |
| S | Solar | The abundance of sulfur given logarithmically with respect to the solar value. |
| Ca | Solar | The abundance of calcium given logarithmically with respect to the solar value. |
| Sc | Solar | The abundance of scandium given logarithmically with respect to the solar value. |
| Ti | Solar | The abundance of titanium given logarithmically with respect to the solar value. |
| V  | Solar | The abundance of vanadium given logarithmically with respect to the solar value. |
| Cr | Solar | The abundance of chromium given logarithmically with respect to the solar value. |
| Mn | Solar | The abundance of manganese given logarithmically with respect to the solar value. |
| Fe | Solar | The abundance of iron given logarithmically with respect to the solar value. |
| Co | Solar | The abundance of cobalt given logarithmically with respect to the solar value. |
| Ni | Solar | The abundance of nickel given logarithmically with respect to the solar value. |
| Cu | Solar | The abundance of copper given logarithmically with respect to the solar value. |
| Zn | Solar | The abundance of zinc given logarithmically with respect to the solar value. |
| Y  | Solar | The abundance of yttrium given logarithmically with respect to the solar value. |
| Zr | Solar | The abundance of zirconium given logarithmically with respect to the solar value. |
| Ba | Solar | The abundance of barium given logarithmically with respect to the solar value. |
| La | Solar | The abundance of lanthanum given logarithmically with respect to the solar value. |
| Ce | Solar | The abundance of cerium given logarithmically with respect to the solar value. |
| Nd | Solar | The abundance of neodymium given logarithmically with respect to the solar value. |
| Sm | Solar | The abundance of samarium given logarithmically with respect to the solar value. |
| Eu | Solar | The abundance of europium given logarithmically with respect to the solar value. |

Table 5
Abundance Details for Z>10

This table has 112 columns of data. The contents are specified below.

| Column | Column Header | Unit | Description |
|---|---|---|---|
| 1 | Star | | Primary Name for the star |
| 2 | Source | | Source of Data / Original Paper - See source key in Table 2 |
| 3 | Key | | Identifier |
| 4 | Phase | | Phase of the observation |
| 5 | Teff | K | Effective Temperature (K) |
| 6 | Log(g) | cm s-2 | Log surface gravity (cm/s^2) |
| 7 | Vt | km s-1 | Microturbulent velocity (km/s) |
| 8 | log ε | | Mean abundance of Na I relative to log ε(Hydrogen) = 12.00 |
| 9 | σ | | Standard deviation of the abundance about the mean |
| 10 | N | | Number of lines used in the mean abundance |
| 11-112 | $...$ | | Columns 11-112 repeat the Na I sequence for Mg I, Al I, Si I, Si II, S I, Ca I, Ca II, Sc I, Sc II, Ti I, Ti II, V I, V II, Cr I, Cr II, Mn I, Fe I, Fe II, Co I, Ni I, Cu I, Zn I, Y I, Y II, Zr I, Zr II, Ba II, La II, Ce II, Pr II, Nd II, Sm II, Eu II |

Table 6

| | | | | | | | | | | Lithium | | Carbon | | | | Nitrogen | Oxygen | | |
|---|---|---|---|---|---|---|---|---|---|---|---|---|---|---|---|---|---|---|---|
| Star | Source | Key | Phase | T | G | Vt | Vb | Fe | Depth | Li | Limit | C2 | 505.20 | 538.00 | 711.00 | <C> | <N> | 616.50 | 630.00 | <O> |
| | | | | (K) | (cms^-2) | (kms^-1) | (kms^-1) | | | | | | | | | | | | | |
| AN Aur | S | anaur | 0.137 | 5689 | 1.53 | 2.99 | 10.20 | 7.27 | 0.00777 | 0.70 | L | | | | | | | 8.52 | 8.58 | 8.57 |
| AP Sgr | S | apsgr | 0.125 | 6310 | 1.98 | 2.81 | 11.00 | 7.60 | 0.00771 | 1.09 | L | | | | | | | 8.82 | 8.38 | 8.71 |
| AS Per | S | asper | 0.592 | 5558 | 1.64 | 2.88 | 9.30 | 7.51 | 0.00475 | 0.60 | L | | | | | | | 8.69 | 8.64 | 8.65 |
| AW Per | S | awpera | 0.779 | 5663 | 1.90 | 3.62 | 14.40 | 7.43 | 0.01941 | 1.22 | L | | | | 8.25 | 8.25 | | 8.24 | 8.78 | 8.64 |
| | S | awperb | 0.937 | 6466 | 2.24 | 3.03 | 13.80 | 7.55 | 0.00475 | 1.37 | L | | | | 8.32 | 8.32 | | 8.61 | 8.49 | 8.61 |
| | S | awperc | 0.097 | 6380 | 1.92 | 2.88 | 12.60 | 7.56 | 0.00699 | 1.00 | L | | | | 8.37 | 8.37 | | 8.71 | 8.73 | 8.72 |
| | S | awperd | 0.238 | 6073 | 1.95 | 2.83 | 11.20 | 7.56 | 0.00790 | 0.96 | L | | | | 8.35 | 8.35 | | 8.66 | 8.75 | 8.68 |
| | S | awpere | 0.394 | 5873 | 1.82 | 2.63 | 11.10 | 7.56 | 0.00500 | 0.90 | L | | | | 8.36 | 8.36 | | 8.59 | 8.74 | 8.70 |
| | S | awperg | 0.674 | 5520 | 1.80 | 3.04 | 12.65 | 7.49 | 0.02638 | 1.37 | | | | | 8.48 | 8.48 | | 8.54 | 8.79 | 8.73 |
| | S | awperh | 0.858 | 6041 | 2.00 | 3.71 | 15.00 | 7.58 | 0.01159 | 1.50 | L | | | | | | | 8.07 | 9.16 | 8.34 |
| | S | awperi | 0.006 | 6597 | 1.86 | 2.96 | 14.20 | 7.51 | 0.00256 | 1.00 | L | | | | | | | 8.73 | 8.37 | 8.73 |
| | S | awperj | 0.629 | 5535 | 1.74 | 2.91 | 11.70 | 7.50 | 0.02618 | 1.28 | | | | | | | | 8.09 | 8.82 | 8.64 |
| | S | awperk | 0.530 | 5603 | 1.57 | 2.82 | 11.30 | 7.53 | 0.02136 | 1.04 | | | | | | | | 8.48 | 8.60 | 8.57 |
| ⋮ | ⋮ | ⋮ | ⋮ | ⋮ | ⋮ | ⋮ | ⋮ | ⋮ | ⋮ | ⋮ | ⋮ | ⋮ | ⋮ | ⋮ | ⋮ | ⋮ | ⋮ | ⋮ | ⋮ | ⋮ |

| Column | Unit | Description |
|---|---|---|
| Star | | Name of Cepheid |
| Source | | Source of Data / Original Paper - See source key in Table 2 |
| Key | | Identifier |
| Phase | | Phase of the observation |
| T | K | Effective Temperature |
| G | cm s-2 | Logarithm of the surface acceleration (gravity) derived from an ionization balance |
| Vt | km s-1 | Microturbulent velocity |
| Vb | km s-1 | Broadening velocity assumed to be macroturbulent profile |
| Iron | log ε | Iron abundance. The solar iron abundance is 7.47 relative to H = 12. |
| Depth | | Measured depth of the Li I 670.7 nm feature |
| Li | log ε | Lithium abundance.  The solar lithium abundance is 1.0 dex |
| Limit | | L = Abundance is an upper limit |
| C2 | log ε | Carbon abundance from C2 Swann lines - primary indicator at 513.5 nm |
| 505.2 | log ε | Carbon abundance from C I 505.2 nm line. |
| 538.0 | log ε | Carbon abundance from C I 538.0 nm line. |
| 711.0 | log ε | Carbon abundance from C I 538.0 nm line. |
| <C> | log ε | Mean carbon abundance - weights discussed in text |
| <N> | log ε | Nitrogen abundance determined from the N I lines at 714.2 and 716.8 nm lines. |
| 615.5 | log ε | Oxygen abundance from O I 615.5 triplet |
| 630.0 | log ε | Oxygen abundance from [O I] 630.0 nm line |
| <O> | log ε | Mean oxygen abundance - weights discussed in text |

Table 7
Average Abundances for Cepheids

This table has 123 columns of data. The contents are specified below.

| Column Number | Column Header | Unit | Description |
|---|---|---|---|
| 1 | Star | | Name of Cepheid |
| 2 | <Fe> | log ε | Average iron abundance |
| 3 | medFe | log ε | Median iron abundance |
| 4 | Range | | Maximum - minimum iron abundance |
| 5 | N | | Number of phases in the average |
| 6 | Li | log ε | Average lithium abundance |
| 7 | Range | | Maximum - minimum iron abundance |
| 8 | N | | Number of phases in the average |
| 9 | <C> | log ε | Average carbon abundance |
| 10 | [C/Fe] | Solar | Average carbon abundance with respect to the Sun normalized to the iron abundance |
| 11 | Range | | Maximum - minimum carbon abundance |
| 12 | N | | Number of phases in the average |
| 13 | <N> | log ε | Average nitrogen abundance |
| 14 | [N/Fe] | Solar | Average nitrogen abundance with respect to the Sun normalized to the iron abundance |
| 15 | Range | | Maximum - minimum nitrogen abundance |
| 16 | N | | Number of phases in the average |
| 17 | <O> | log ε | Average oxygen abundance |
| 18 | [O/Fe] | Solar | Average oxygen abundance with respect to the Sun normalized to the iron abundance |
| 19 | Range | | Maximum - minimum oxygen abundance |
| 20 | N | | Number of phases in the average |
| 21 | <[Na/H]> | Solar | Average sodium abundance |
| 22 | [Na/Fe] | Solar | Average sodium abundance with respect to the Sun normalized to the iron abundance |

| | | |
|---|---|---|
| 23 | Range | Maximum - minimum sodium abundance |
| 24 | N | Number of phases in the average |
| 25-123 | $...$ | Columns 25-123 repeat the sodium sequence for Mg, Al, Si, Si, S, Ca, Ca, Sc, Sc, Ti, Ti, V, Cr, Mn, Fe, Co, Ni, Cu, Zn, Y, Zr, Ba, La, Ce, Pr, Nd, Sm, Eu |

These mean abundances are normalized to the solar values of Luck (2018).

Table 8
Internal Mean Uncertainties

| | σ(T) | | | | | |
|---|---|---|---|---|---|---|
| | Combined | S | H | M | F1 | F2 |
| Mean | 77 | 67 | 101 | 62 | 60 | 87 |
| Median | 65 | 61 | 86 | 58 | 57 | 70 |
| STD DEV | 46 | 33 | 60 | 21 | 27 | 54 |
| N | 1127 | 603 | 315 | 30 | 102 | 77 |
| Min | 21 | 21 | 21 | 27 | 25 | 22 |
| Max | 398 | 294 | 398 | 106 | 138 | 284 |

| | σ(Fe I) | | | | | |
|---|---|---|---|---|---|---|
| | Combined | S | H | M | F1 | F2 |
| Mean | 0.14 | 0.13 | 0.16 | 0.15 | 0.14 | 0.17 |
| Median | 0.14 | 0.12 | 0.15 | 0.15 | 0.14 | 0.15 |
| STD DEV | 0.04 | 0.03 | 0.05 | 0.01 | 0.02 | 0.05 |
| N | 1127 | 603 | 315 | 30 | 107 | 77 |
| Min | 0.09 | 0.09 | 0.11 | 0.13 | 0.11 | 0.12 |
| Max | 0.44 | 0.28 | 0.44 | 0.19 | 0.24 | 0.39 |

S, H, M, F1, F2 refer to the datasets subsets as given in Tables 3 - 6.
σ(T) is the standard deviation of the temperature determination
σ(Fe I) is the standard deviation of the Fe I per line abundances about the mean value

Table 9
Parameter Sensitivity

| | Ensemble log ε | | | | Delta WRT 5757/1.73 | | |
|---|---|---|---|---|---|---|---|
| Fe I | 1.43 | 1.73 | 2.03 | Fe I | 1.43 | 1.73 | 2.03 |
| 5657 | 7.560 | 7.554 | 7.548 | | -0.069 | -0.075 | -0.081 |
| 5757 | 7.636 | 7.629 | 7.622 | | 0.007 | 0 | -0.007 |
| 5857 | 7.709 | 7.701 | 7.694 | | 0.080 | 0.072 | 0.065 |
| | | | | | | | |
| Fe II | 1.43 | 1.73 | 2.03 | Fe II | 1.43 | 1.73 | 2.03 |
| 5657 | 7.538 | 7.650 | 7.765 | | -0.090 | 0.022 | 0.137 |
| 5757 | 7.519 | 7.628 | 7.740 | | -0.109 | 0 | 0.112 |
| 5857 | 7.505 | 7.611 | 7.720 | | -0.123 | -0.017 | 0.092 |
| | | | | | | | |
| | | Fe I-Fe II | 1.43 | 1.73 | 2.03 | | |
| | | 5657 | 0.022 | -0.096 | -0.217 | | |
| | | 5757 | 0.117 | 0.001 | -0.118 | | |
| | | 5857 | 0.204 | 0.09 | -0.026 | | |
| | | | | | | | |
| | | Vt | 2.77 | 3.07 | 3.37 | | |
| | | Fe I | 7.674 | 7.629 | 7.591 | | |
| | | Fe II | 7.683 | 7.628 | 7.582 | | |

These are computed using the ysgrf iron ensemble.

The Fe I - Fe II block shows the difference in iron abundance between Fe I and Fe II as a function of parameter change.

Table 10
Parameter and Iron Abundance Differences

|     |        | Combined | S     | H     | M     | F1    | F2    |
|-----|--------|----------|-------|-------|-------|-------|-------|
| dT  | Mean   | 6        | 11    | 11    | 2     | 1     | -37   |
|     | Median | 3        | 11    | 2     | 2     | 1     | -24   |
|     | σ      | 53       | 47    | 57    | 30    | 11    | 87    |
|     | N      | 1046     | 524   | 315   | 30    | 102   | 75    |
|     | Min    | -359     | -301  | -90   | -128  | -64   | -359  |
|     | Max    | 662      | 275   | 662   | 70    | 42    | 229   |
| dG  | Mean   | -0.09    | -0.07 | -0.11 | -0.11 | -0.24 | 0.05  |
|     | Median | -0.10    | -0.07 | -0.12 | -0.13 | -0.26 | 0.11  |
|     | σ      | 0.23     | 0.21  | 0.26  | 0.19  | 0.17  | 0.27  |
|     | N      | 1046     | 524   | 315   | 30    | 102   | 75    |
|     | Min    | -1.32    | -1.28 | -1.06 | -0.41 | -0.56 | -1.32 |
|     | Max    | 1.24     | 0.91  | 1.24  | 0.58  | 0.65  | 0.58  |
| dV  | Mean   | -0.89    | -0.94 | -0.90 | -1.00 | -1.20 | 0.00  |
|     | Median | -0.82    | -0.87 | -0.74 | -0.85 | -1.12 | 0.01  |
|     | σ      | 0.67     | 0.60  | 0.70  | 0.58  | 0.68  | 0.23  |
|     | N      | 1046     | 524   | 315   | 30    | 102   | 75    |
|     | Min    | -4.45    | -3.16 | -4.45 | -2.83 | -3.18 | -1.04 |
|     | Max    | 2.43     | 2.43  | 0.16  | -0.41 | 1.83  | 0.84  |
| dFe | Mean   | -0.03    | 0.00  | -0.05 | -0.06 | -0.07 | -0.02 |
|     | Median | -0.03    | -0.01 | -0.06 | -0.06 | -0.07 | -0.02 |
|     | σ      | 0.07     | 0.07  | 0.07  | 0.03  | 0.04  | 0.06  |
|     | N      | 1046     | 524   | 315   | 30    | 102   | 75    |
|     | Min    | -0.36    | -0.20 | -0.36 | -0.11 | -0.13 | -0.17 |
|     | Max    | 0.51     | 0.33  | 0.51  | 0.01  | 0.13  | 0.11  |
|     |        | L13      | G     | K     |       | G-L13 |       |
| dFe | Mean   | 0.10     | -0.02 | 0.04  |       | 0.11  |       |
|     | Median | 0.10     | -0.03 | 0.04  |       | 0.09  |       |
|     | σ      | 0.10     | 0.08  | 0.05  |       | 0.13  |       |
|     | N      | 62       | 57    | 17    |       | 25    |       |
|     | Min    | -0.08    | -0.22 | -0.03 |       | -0.16 |       |
|     | Max    | 0.60     | 0.16  | 0.15  |       | 0.36  |       |

Differences are (this work – reference) except for G-L13

| | | | |
|---|---|---|---|
| dT | Effective temperature difference | dV | Microturbulent velocity difference |
| dG | Gravity difference | dFe | Iron abundance difference |

| Analyses | Reference |
|---|---|
| S | Andrievsky et al. 2002a,b,c |
|   | Luck et al. 2003 |
|   | Andrievsky et al. 2004 |
|   | Luck & Andrievsky 2004 |
|   | Kovtyukh et al. 2005 |
|   | Andrievsky et al. 2005 |
|   | Luck et al. 2006 |
|   | Luck et al. 2008 |
| H | Luck & Lambert 2011 |
| M | Luck & Lambert 2011 |
| F1 | Luck & Lambert 2011 |
| F2 | Luck 2014 |
| Combined | Over S,H,M,F1, and F2 |
| L13 | Lemasle et al 2013 |
| G | Genovali et al 2014, 2015 |
| K | Kovtyukh et al 2016 |

Table 11
Average Parameters for Cepheids

| Cepheid | Period (days) | N | <T> (K) | <G> (cm s^-2) | <Vt> (km s^-1) | <Vb> (km s^-1) | <Fe> (log ε) | Range (log ε) |
|---|---|---|---|---|---|---|---|---|
| V473 Lyr | 1.49078 | 5 | 6019 | 2.30 | 3.55 | 7.98 | 7.37 | 0.05 |
| SU Cas | 1.949324 | 13 | 6274 | 2.26 | 2.41 | 9.80 | 7.44 | 0.08 |
| DT Cyg | 2.499215 | 14 | 6192 | 2.27 | 2.63 | 9.85 | 7.52 | 0.09 |
| SZ Tau | 3.14873 | 16 | 5987 | 2.03 | 2.86 | 11.39 | 7.49 | 0.09 |
| V1334 Cyg | 3.332816 | 11 | 6293 | 2.22 | 2.81 | 16.13 | 7.51 | 0.09 |
| RT Aur | 3.728485 | 12 | 5948 | 2.06 | 2.75 | 8.70 | 7.54 | 0.07 |
| SU Cyg | 3.845547 | 12 | 6036 | 2.08 | 3.21 | 15.97 | 7.41 | 0.20 |
| ST Tau | 4.034299 | 7 | 6052 | 2.07 | 3.04 | 9.00 | 7.41 | 0.04 |
| BQ Ser | 4.2709 | 7 | 6040 | 2.16 | 3.26 | 15.08 | 7.49 | 0.09 |
| Y Lac | 4.323776 | 10 | 5915 | 1.87 | 3.62 | 15.37 | 7.42 | 0.19 |
| T Vul | 4.435462 | 12 | 5852 | 2.03 | 3.20 | 10.82 | 7.43 | 0.11 |
| FF Aql | 4.470881 | 14 | 6164 | 2.04 | 3.15 | 10.41 | 7.52 | 0.07 |
| CF Cas | 4.87522 | 7 | 5672 | 1.74 | 3.41 | 10.60 | 7.45 | 0.14 |
| BG Lac | 5.331908 | 9 | 5674 | 1.70 | 3.20 | 10.35 | 7.46 | 0.07 |
| del Cep | 5.366341 | 19 | 5854 | 1.96 | 3.00 | 9.69 | 7.55 | 0.05 |
| Y Sgr | 5.77335 | 12 | 5767 | 1.77 | 3.53 | 15.44 | 7.52 | 0.26 |
| FM Aql | 6.11429 | 12 | 5766 | 1.68 | 3.41 | 16.28 | 7.56 | 0.20 |
| X Vul | 6.319588 | 8 | 5753 | 1.81 | 3.12 | 9.12 | 7.55 | 0.06 |
| XX Sgr | 6.42414 | 5 | 5805 | 1.81 | 3.05 | 11.44 | 7.54 | 0.07 |
| AW Per | 6.463589 | 11 | 5928 | 1.86 | 3.00 | 12.47 | 7.52 | 0.15 |
| U Sgr | 6.745226 | 11 | 5709 | 1.79 | 3.39 | 12.61 | 7.54 | 0.14 |
| U Aql | 7.024049 | 5 | 5565 | 1.64 | 3.26 | 9.04 | 7.59 | 0.03 |
| eta Aql | 7.176915 | 14 | 5746 | 1.86 | 3.30 | 12.45 | 7.56 | 0.09 |
| BB Her | 7.507945 | 8 | 5641 | 1.65 | 3.33 | 11.42 | 7.66 | 0.04 |
| RS Ori | 7.566881 | 7 | 5891 | 1.77 | 2.99 | 13.27 | 7.46 | 0.10 |
| V440 Per | 7.57 | 10 | 6056 | 1.97 | 3.00 | 7.98 | 7.37 | 0.05 |
| W Sgr | 7.59503 | 9 | 5765 | 1.78 | 3.12 | 10.10 | 7.47 | 0.12 |
| RX Cam | 7.912024 | 10 | 5703 | 1.65 | 3.20 | 11.34 | 7.49 | 0.10 |
| W Gem | 7.913779 | 13 | 5771 | 1.69 | 3.25 | 9.72 | 7.43 | 0.10 |
| U Vul | 7.990676 | 8 | 5779 | 1.73 | 3.11 | 9.54 | 7.56 | 0.05 |
| DL Cas | 8.000669 | 11 | 5682 | 1.56 | 3.57 | 16.01 | 7.45 | 0.22 |
| V636 Cas | 8.37571 | 8 | 5505 | 1.47 | 3.20 | 10.08 | 7.55 | 0.03 |
| S Sge | 8.382086 | 11 | 5689 | 1.73 | 3.22 | 10.66 | 7.56 | 0.07 |
| V500 Sco | 9.316863 | 5 | 5675 | 1.56 | 3.11 | 9.06 | 7.45 | 0.03 |
| FN Aql | 9.48164 | 11 | 5488 | 1.38 | 3.16 | 10.42 | 7.39 | 0.07 |
| YZ Sgr | 9.553606 | 11 | 5653 | 1.69 | 3.29 | 10.57 | 7.56 | 0.09 |
| zet Gem | 10.15073 | 12 | 5512 | 1.52 | 3.12 | 9.68 | 7.52 | 0.08 |
| Z Lac | 10.885613 | 10 | 5618 | 1.49 | 3.41 | 11.40 | 7.50 | 0.12 |
| VX Per | 10.88904 | 12 | 5783 | 1.64 | 3.23 | 9.82 | 7.43 | 0.06 |
| RX Aur | 11.626 | 13 | 5782 | 1.67 | 3.51 | 15.39 | 7.46 | 0.16 |

| Cepheid | Period | N | <T> | <G> | <Vt> | <Vb> | <Fe> | Range |
|---|---|---|---|---|---|---|---|---|
| TT Aql | 13.754912 | 10 | 5272 | 1.15 | 3.50 | 11.83 | 7.61 | 0.15 |
| SV Mon | 15.23278 | 9 | 5330 | 1.11 | 3.33 | 9.46 | 7.48 | 0.10 |
| X Cyg | 16.386332 | 20 | 5252 | 1.10 | 3.62 | 12.02 | 7.56 | 0.32 |
| RW Cam | 16.415014 | 17 | 5213 | 1.03 | 3.29 | 11.91 | 7.56 | 0.18 |
| CD Cyg | 17.073967 | 17 | 5394 | 1.19 | 3.55 | 11.62 | 7.63 | 0.16 |
| Y Oph | 17.12413 | 14 | 5819 | 1.62 | 3.24 | 9.19 | 7.50 | 0.05 |
| SZ Aql | 17.141247 | 11 | 5398 | 1.20 | 3.71 | 12.46 | 7.66 | 0.20 |
| WZ Sgr | 21.849709 | 10 | 5140 | 0.88 | 3.86 | 11.20 | 7.75 | 0.32 |
| X Pup | 25.961 | 8 | 5353 | 0.75 | 3.54 | 11.99 | 7.48 | 0.06 |
| T Mon | 27.024649 | 12 | 5108 | 0.93 | 3.72 | 11.96 | 7.68 | 0.18 |
| SV Vul | 45.0121 | 15 | 5329 | 0.85 | 4.06 | 12.80 | 7.58 | 0.14 |
| S Vul | 68.463997 | 6 | 5452 | 0.93 | 4.79 | 14.22 | 7.56 | 0.10 |

| Column | Unit | Description |
|---|---|---|
| Cepheid | | Cephied name |
| Period | days | Period |
| N | | Number of phases observed |
| <T> | K | Average effective temperature over cycle |
| <G> | cm s^-2 | Average gravity over cycle |
| <Vt> | km s^-1 | Average microturbulent velocity over cycle |
| <Vb> | km s^-1 | Average broadening velocity over cycle |
| <Fe> | log ε | Average total iron abundance |
| Range | log ε | Range in determined iron abundance |

Table 12

Radial Abundance Gradients: d[x/X] = a * $R_g$ + b  dex/kpc

|        | a       | b       | e_a    | e_b    | σ      |        | a       | b       | e_a    | e_b    | σ      | N   |
|--------|---------|---------|--------|--------|--------|--------|---------|---------|--------|--------|--------|-----|
| [C/H]  | -0.0665 |  0.3522 | 0.0038 | 0.0351 | 0.1633 | [C/Fe] | -0.0150 | -0.1174 | 0.0033 | 0.0304 | 0.1412 | 368 |
| [N/H]  | -0.0470 |  0.8278 | 0.0040 | 0.0378 | 0.1700 | [N/Fe] |  0.0029 |  0.3769 | 0.0036 | 0.0340 | 0.1531 | 342 |
| [O/H]  | -0.0429 |  0.3839 | 0.0023 | 0.0210 | 0.1026 | [O/Fe] |  0.0077 | -0.0738 | 0.0024 | 0.0218 | 0.1065 | 421 |
| [Na/H] | -0.0517 |  0.8167 | 0.0034 | 0.0314 | 0.1544 | [Na/Fe]| -0.0008 |  0.3576 | 0.0026 | 0.0237 | 0.1166 | 422 |
| [Mg/H] | -0.0574 |  0.6022 | 0.0035 | 0.0327 | 0.1608 | [Mg/Fe]| -0.0066 |  0.1431 | 0.0026 | 0.0241 | 0.1185 | 422 |
| [Al/H] | -0.0550 |  0.6040 | 0.0033 | 0.0303 | 0.1490 | [Al/Fe]| -0.0042 |  0.1448 | 0.0023 | 0.0209 | 0.1031 | 422 |
| [Si/H] | -0.0497 |  0.6023 | 0.0022 | 0.0201 | 0.0992 | [Si/Fe]|  0.0011 |  0.1431 | 0.0008 | 0.0074 | 0.0364 | 422 |
| [S/H]  | -0.0693 |  0.6612 | 0.0035 | 0.0327 | 0.1610 | [S/Fe] | -0.0184 |  0.2021 | 0.0023 | 0.0214 | 0.1056 | 422 |
| [Ca/H] | -0.0511 |  0.5049 | 0.0026 | 0.0245 | 0.1206 | [Ca/Fe]| -0.0002 |  0.0458 | 0.0013 | 0.0121 | 0.0594 | 422 |
| [Sc/H] | -0.0438 |  0.7254 | 0.0033 | 0.0303 | 0.1492 | [Sc/Fe]|  0.0070 |  0.2662 | 0.0025 | 0.0228 | 0.1122 | 422 |
| [Ti/H] | -0.0365 |  0.4566 | 0.0025 | 0.0228 | 0.1121 | [Ti/Fe]|  0.0144 | -0.0025 | 0.0014 | 0.0130 | 0.0642 | 422 |
| [V/H]  | -0.0305 |  0.3922 | 0.0029 | 0.0267 | 0.1317 | [V/Fe] |  0.0203 | -0.0669 | 0.0025 | 0.0230 | 0.1133 | 422 |
| [Cr/H] | -0.0444 |  0.5559 | 0.0026 | 0.0241 | 0.1188 | [Cr/Fe]|  0.0064 |  0.0968 | 0.0013 | 0.0122 | 0.0603 | 422 |
| [Mn/H] | -0.0489 |  0.3161 | 0.0031 | 0.0286 | 0.1409 | [Mn/Fe]|  0.0019 | -0.1431 | 0.0017 | 0.0159 | 0.0781 | 422 |
| [Fe/H] | -0.0508 |  0.4591 | 0.0022 | 0.0203 | 0.1000 | …      | …       | …       | …      | …      | …      | 422 |
| [Co/H] | -0.0218 |  0.4024 | 0.0034 | 0.0317 | 0.1562 | [Co/Fe]|  0.0290 | -0.0568 | 0.0033 | 0.0303 | 0.1492 | 422 |
| [Ni/H] | -0.0532 |  0.4594 | 0.0026 | 0.0237 | 0.1167 | [Ni/Fe]| -0.0024 |  0.0003 | 0.0008 | 0.0075 | 0.0368 | 422 |
| [Cu/H] | -0.0536 |  0.6160 | 0.0051 | 0.0466 | 0.2222 | [Cu/Fe]| -0.0049 |  0.1772 | 0.0047 | 0.0436 | 0.2079 | 414 |
| [Zn/H] | -0.0728 |  0.5526 | 0.0053 | 0.0491 | 0.2380 | [Zn/Fe]| -0.0234 |  0.1101 | 0.0046 | 0.0429 | 0.2079 | 412 |
| [Y/H]  | -0.0302 |  0.4774 | 0.0031 | 0.0283 | 0.1395 | [Y/Fe] |  0.0206 |  0.0182 | 0.0022 | 0.0205 | 0.1008 | 422 |
| [Zr/H] | -0.0042 |  0.4656 | 0.0049 | 0.0458 | 0.2253 | [Zr/Fe]|  0.0466 |  0.0065 | 0.0050 | 0.0466 | 0.2294 | 422 |
| [Ba/H] | -0.0761 |  0.9254 | 0.0140 | 0.1416 | 0.2736 | [Ba/Fe]| -0.0226 |  0.4991 | 0.0124 | 0.1251 | 0.2419 |  68 |
| [La/H] | -0.0227 |  0.3844 | 0.0031 | 0.0283 | 0.1391 | [La/Fe]|  0.0281 | -0.0747 | 0.0026 | 0.0243 | 0.1195 | 422 |
| [Ce/H] | -0.0129 |  0.2042 | 0.0035 | 0.0325 | 0.1598 | [Ce/Fe]|  0.0379 | -0.2549 | 0.0032 | 0.0296 | 0.1459 | 422 |
| [Nd/H] | -0.0170 |  0.1683 | 0.0029 | 0.0267 | 0.1316 | [Nd/Fe]|  0.0338 | -0.2909 | 0.0022 | 0.0207 | 0.1019 | 422 |
| [Sm/H] | -0.0236 |  0.1843 | 0.0035 | 0.0326 | 0.1587 | [Sm/Fe]|  0.0280 | -0.2831 | 0.0030 | 0.0280 | 0.1362 | 404 |
| [Eu/H] | -0.0226 | -0.0557 | 0.0028 | 0.0257 | 0.1263 | [Eu/Fe]|  0.0283 | -0.5148 | 0.0022 | 0.0207 | 0.1017 | 422 |

$R_G$ computed using the Bayesian distances from Bailer-Jones et al. 2018.

Figures

Figure 1:    Top Panel: Per phase gravities in Cepheids determined from the ionization balance versus the mean gravity found using period-radius-mass relations. Bottom panel: The mean gravity averaged over phase for 52 Cepheids with multiphase gravity determinations versus the gravity determined from period-radius-mass relations. The gravities are in good accord.

Figure 2:    Syntheses of $C_2$ in three Cepheids. There are four syntheses in each panel, the best fitting abundance, this abundance ±0.1 dex, and a synthesis with no $C_2$.

Figure 3:    The C I line at 505.2 nm in three Cepheids. There are four syntheses in each panel, the best fitting abundance, this abundance ±0.1 dex, and a synthesis with no carbon line.

Figure 4:    Syntheses of the C I line at 538.0 nm in three Cepheids. There are four syntheses in each panel, the best fitting abundance, this abundance ±0.1 dex, and a synthesis with no carbon line.

Figure 5:    The C I lines at 711.3 nm in three Cepheids. There are four syntheses in each panel, the best fitting abundance, this abundance ±0.1 dex, and a synthesis with no carbon lines.

Figure 6:    Syntheses of the N I lines at 714.2 and 716.8 nm in three Cepheids. There are four syntheses in each panel, the best fitting abundance, this abundance ±0.1 dex, and a synthesis with no nitrogen lines.

Figure 7:    The O I triplet at 615.6 nm in three Cepheids. There are four syntheses in each panel, the best fitting abundance, this abundance ±0.1 dex, and a synthesis with no oxygen lines.

Figure 8:    Syntheses of the [O I] line at 630.0 nm in three Cepheids. There are four syntheses in each panel, the best fitting abundance, this abundance ±0.1 dex, and a synthesis with no oxygen line.

Figure 9:    The top panel (panel a) shows the GAIA DR2 parallax distances versus the PL derived distances. The error bars for the GAIA data are directly from the uncertainty in the parallax. The error bars for the PL distances assume a total uncertainty of ±0.25 magnitudes. As expected, when the DR2 parallaxes grow smaller, the distance uncertainties increase. In the bottom panel (b), the DR2 data is restricted to parallaxes exceeding eight times their uncertainty. It appears that the PL distances are overall too small.

Figure 10:    The top panel (panel a) shows the Bayesian distance estimate (Bailer-Jones et al. 2018) versus the GAIA DR2 parallax distance. The Bayesian distances deviate systematically from the raw parallax values. Panel (b) shows the Bayesian distance estimate versus the PL distance for the 420 Cepheids with PL distances in this work. The scales are better aligned than the PL versus raw GAIA DR2 comparison.

Figure 11:    Multiphase results for SZ Tau. The top to bottom panels are effective temperature, log (g), microturbulent velocity, total broadening velocity, and log $\varepsilon_{Fe}$. log $\varepsilon_{Fe}$ is the total iron abundance relative to log $\varepsilon_H = 12$. The error bars shown are typical uncertainties derived from the total dataset. SZ Tau is an s-Cepheid (DCEPS) with a small light amplitude. This is reflected in the amplitude of the effective temperature and lack of a clear pulsation signal in its gravity. The label in the bottom right of each panel gives the spectrum source.

Figure 12: Multiphase results for δ Cep. The pulsation signal is very evident in all atmosphere parameter variables. The total range in [Fe/H] is 0.052 dex, much less than the typical uncertainty in a single [Fe/H] determination. The variables, error bars, and source labels are the same as in Figure 11.

Figure 13: Multiphase results for η Aql. The variables, error bars, and source labels are the same as in Figure 11.

Figure 14: Multiphase results for ζ Gem. The variables, error bars, and source labels are the same as in Figure 11. ζ Gem is considered an s-Cepheid (Luck et al. 2008) accounting for its lower temperature amplitude.

Figure 15: Multiphase results for X Cyg. The variables, error bars, and source labels are the same as in Figure 11. Note the correspondence between the microturbulent and total broadening velocity curves.

Figure 16: Multiphase results for T Mon. The variables, error bars, and source labels are the same as in Figure 11.

Figure 17: Multiphase results for SV Vul. The variables, error bars, and source labels are the same as in Figure 11.

Figure 18: Average atmospheric parameter data as a function of log (Period) for the 52 Cepheids having more than five phase points available. The panels show (top to bottom) effective temperature, gravity, microturbulent velocity, total broadening velocity, and $< \log \varepsilon_{Fe} >$. The error bars for the atmospheric parameters are typical values for an individual phase determination of the respective quantity. The behavior of each quantity is systematic. However, there is a cluster of Cepheids with high total broadening relative to the bulk of the stars. The $< \log \varepsilon_{Fe} >$ data indicates no systematic behavior relative to the period in the well-determined abundances. The error bars for $< \log \varepsilon Fe >$ show the total range about the mean for iron.

Figure 19. Cepheid [x/Fe] ratios as a function of element. The features of primary interest are the lowered carbon abundance, enhanced nitrogen content, and a sodium overabundance. See text for a more extended discussion.

Figure 20. The [O/Fe], [C/Fe], and [N/Fe] ratios versus [Fe/H] for Cepheid variables. [O/Fe] shows a distinct dependence on [Fe/H] whereas the [C/Fe] and [N/Fe] ratios do not. The error bars shown are representative of a single determination.

Figure 21: The [O/Na] ratio as a function of log(Period). If the Na overabundance noted in Figure 19 were a result of the NaNe cycle, the expected result would be a dependence of [O/Na] of mass and thus period. There is no dependence evident in the data. The error bars shown are representative of a single determination.

Figure 22. The iron gradient in the Milky Way as determined from Cepheids. The distances in the panel (a) are those from the raw GAIA DR2 parallaxes. In panel (b), the distances are the Bayesian estimates from Bailer-Jones for the same stars. The raw parallaxes show a significant degree of uncertainty at Galactocentric distances larger than about 15 kpc while the "corrected" values are much tighter. The apparent flattening of the gradient at RG > 15 kpc in the DR2 result is not present is the "corrected" distances. This leads to the larger gradient found using the "corrected" distances. The cyan "fit" in each plot is a LOWESS smoothing of the data. The smoothed data exhibits a behavior much like the linear fit except inward of the solar region (R0= 7.9 kpc) where an upturn is indicated.

Figure 23. The iron gradient in the Milky Way as determined from Cepheids. The distances in the panel (a) are those from the raw GAIA DR2 parallaxes limited to those with parallaxes greater than five times their uncertainty. Relative to the unfiltered parallaxes (Figure 22a), the data is cleaner, but the gradient is not significantly different. In the bottom panel, the iron data is shown versus the period-luminosity derived distances. This data is relatively clean, and returns a gradient much like that determined in previous determinations: here the gradient is -0.05 dex/kpc while older work finds -0.06 dex/kpc. The cyan line is as in Figure 22.

Figure 24. The gradients for all species considered versus species. Gradients for Fe-peak and lighter elements have similar magnitude and heavier elements have shallower gradients than the lighter elements.

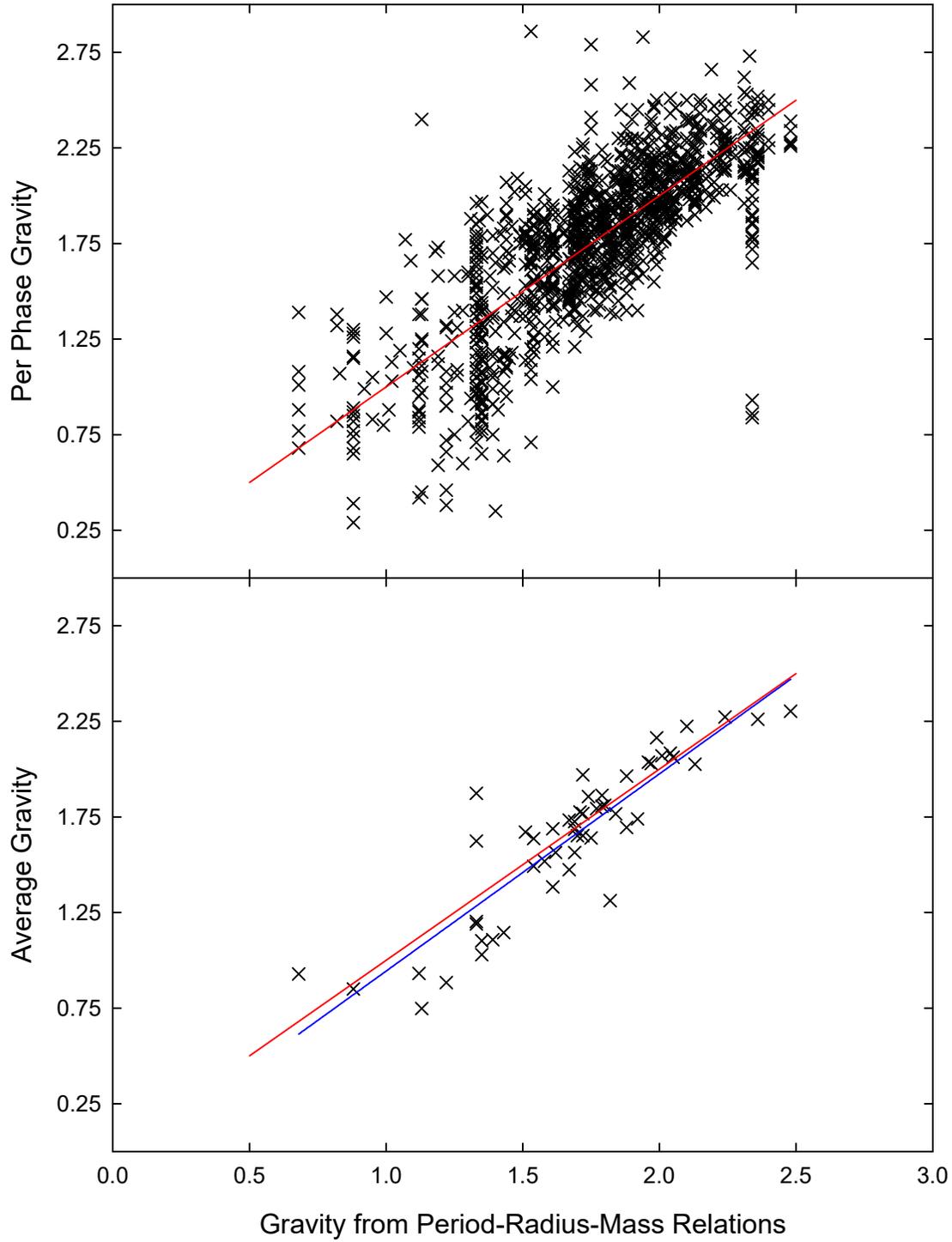

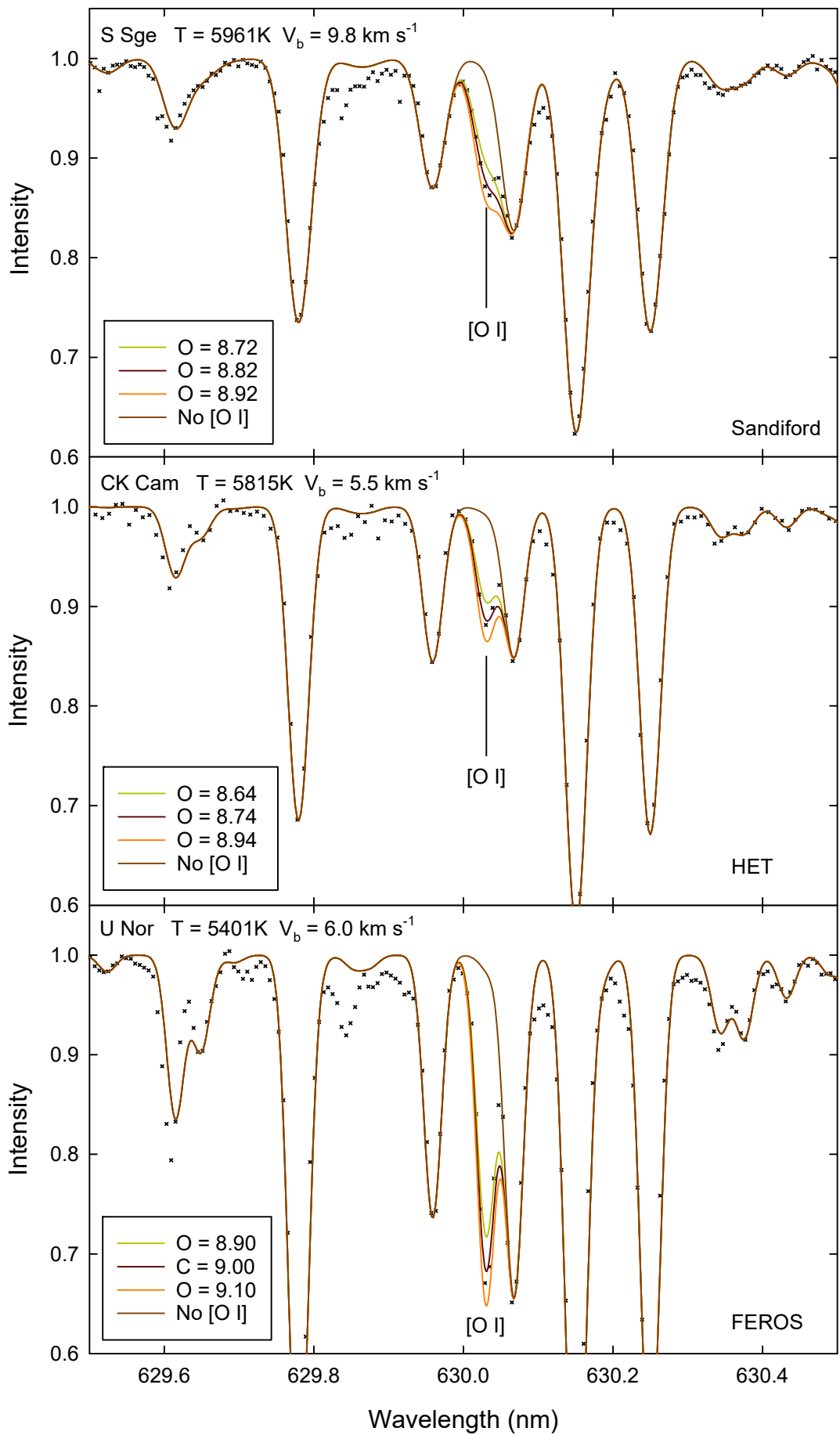

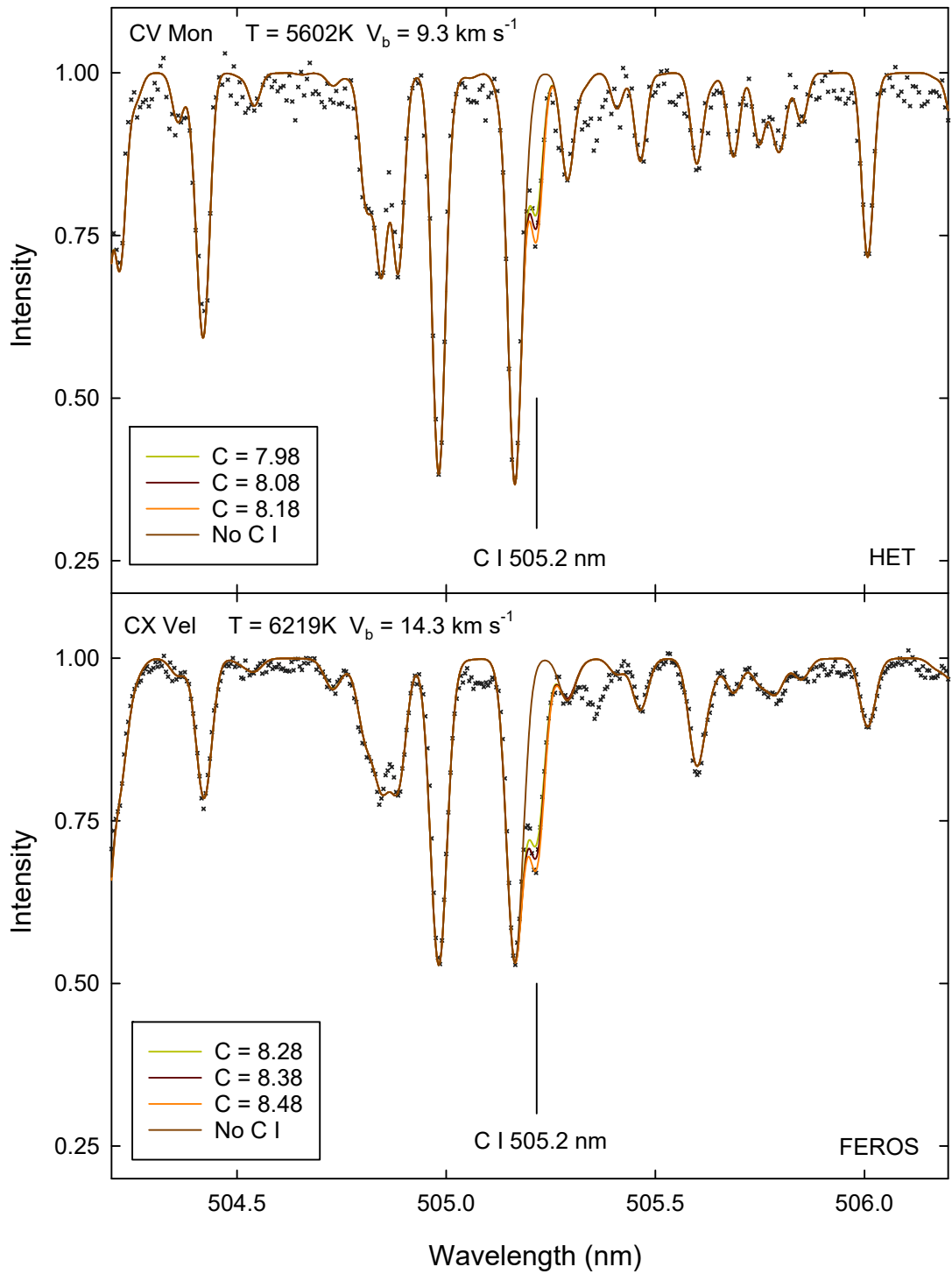

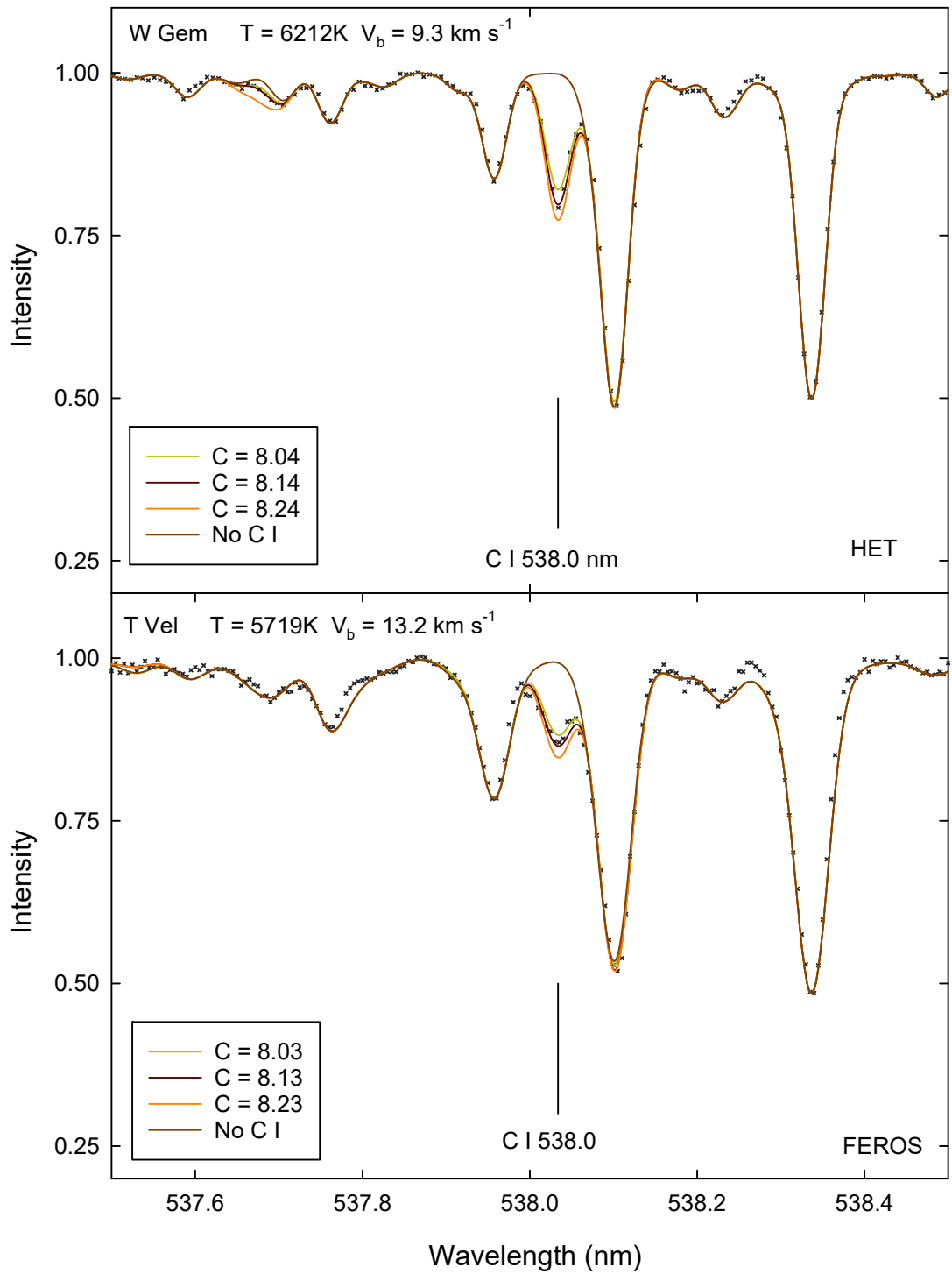

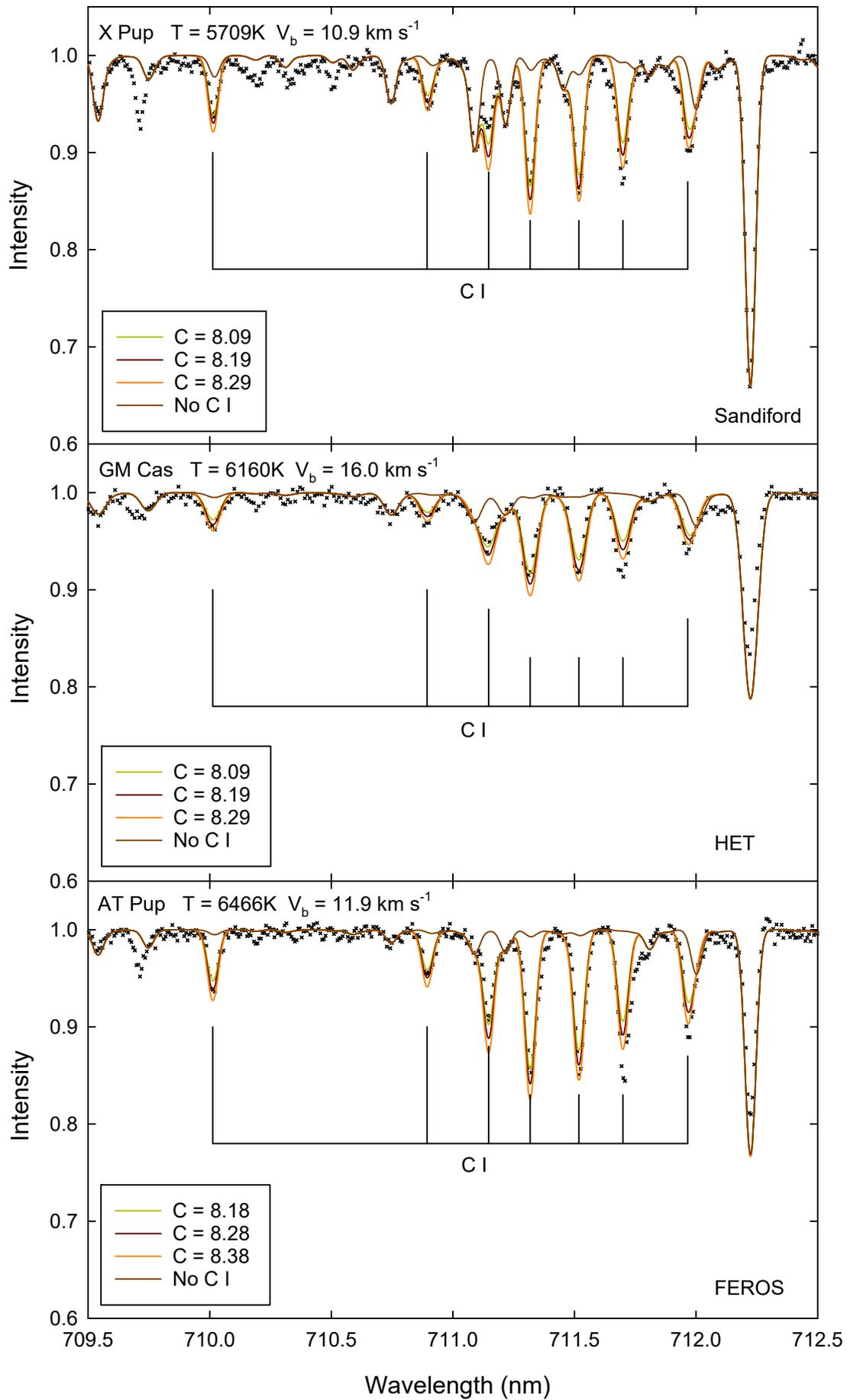

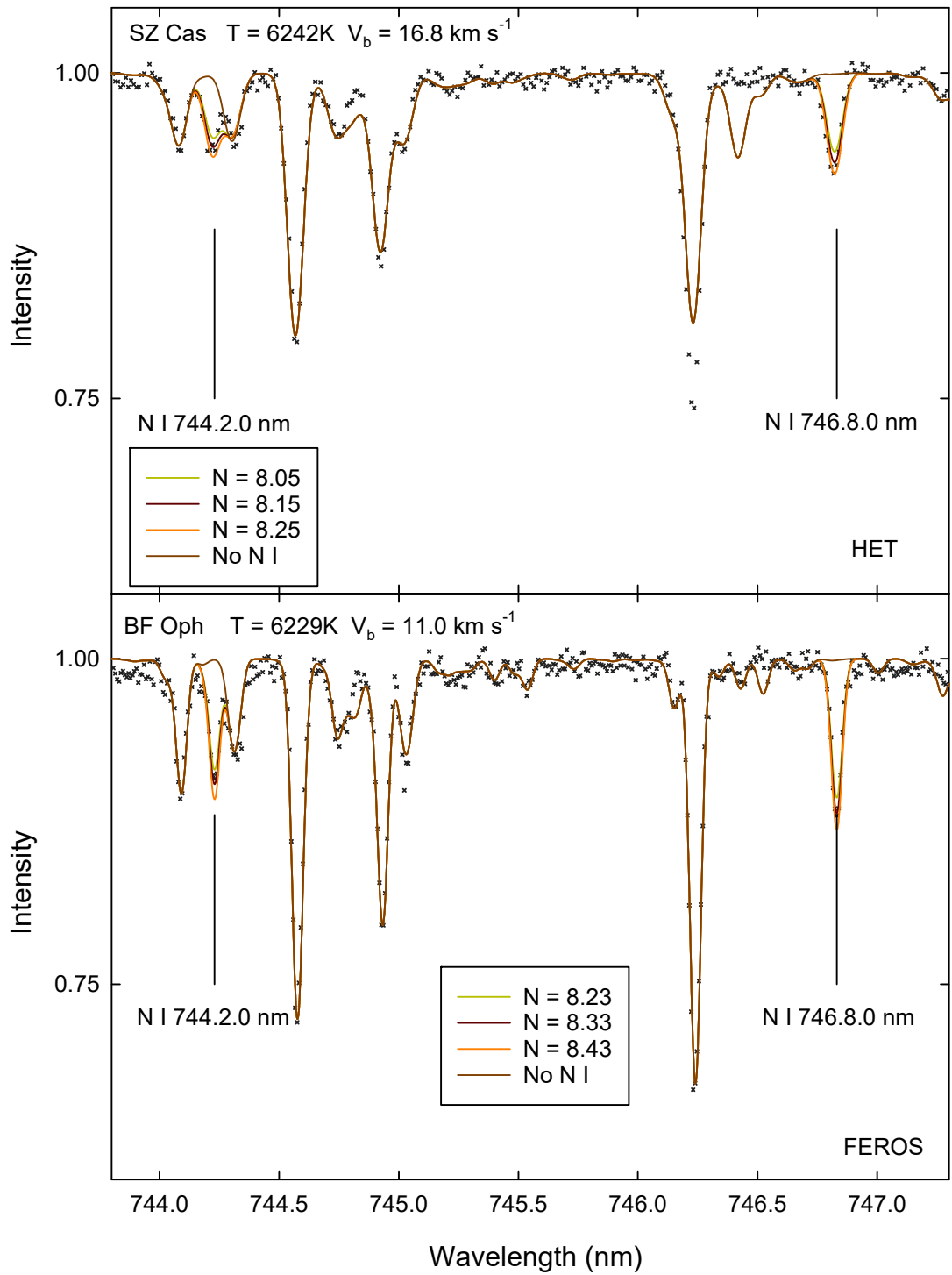

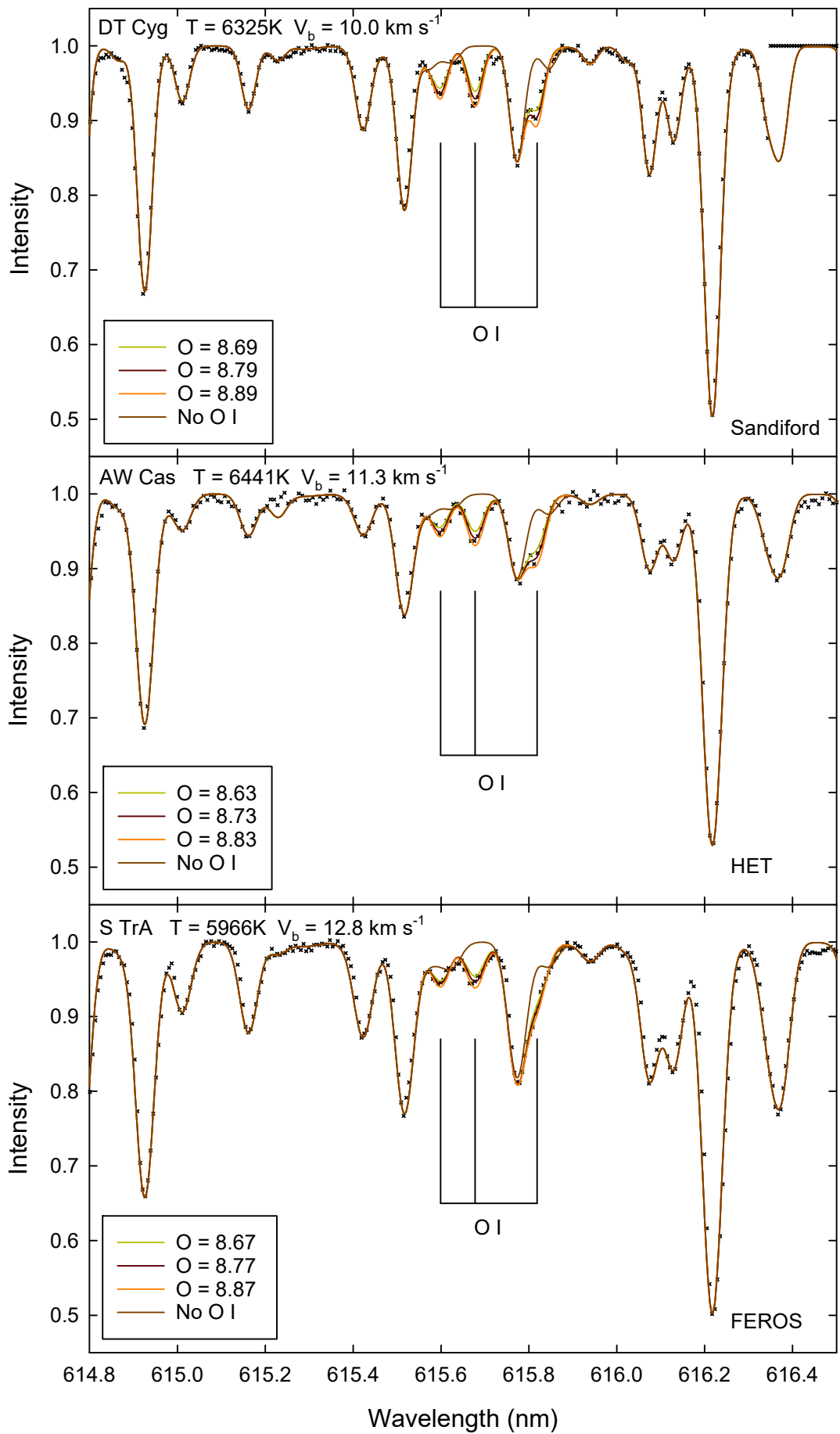

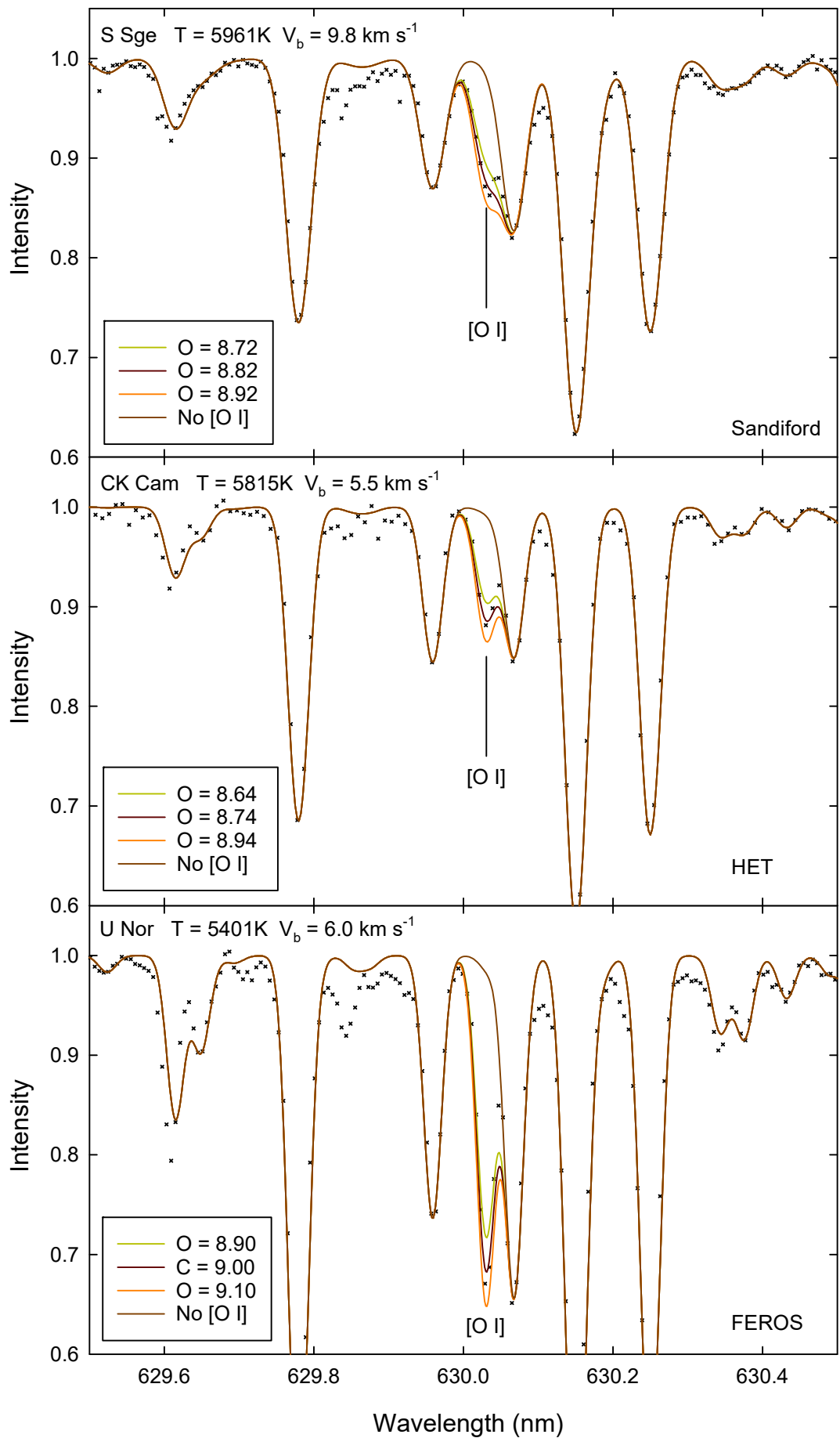

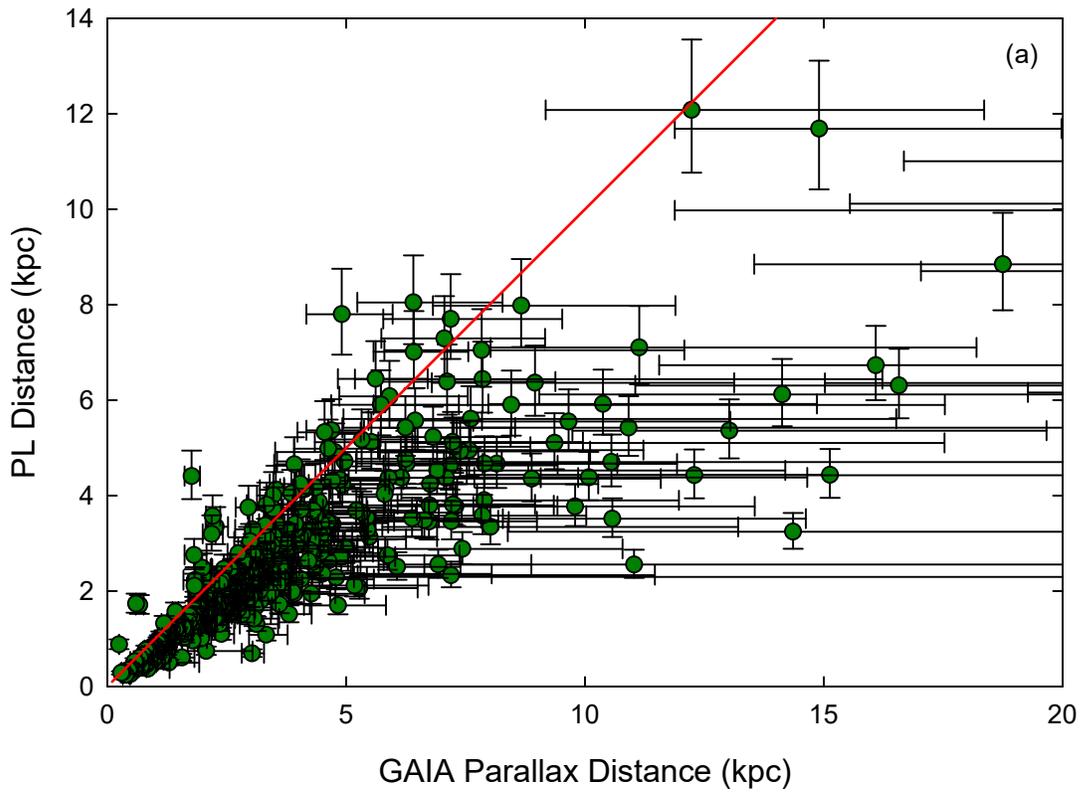

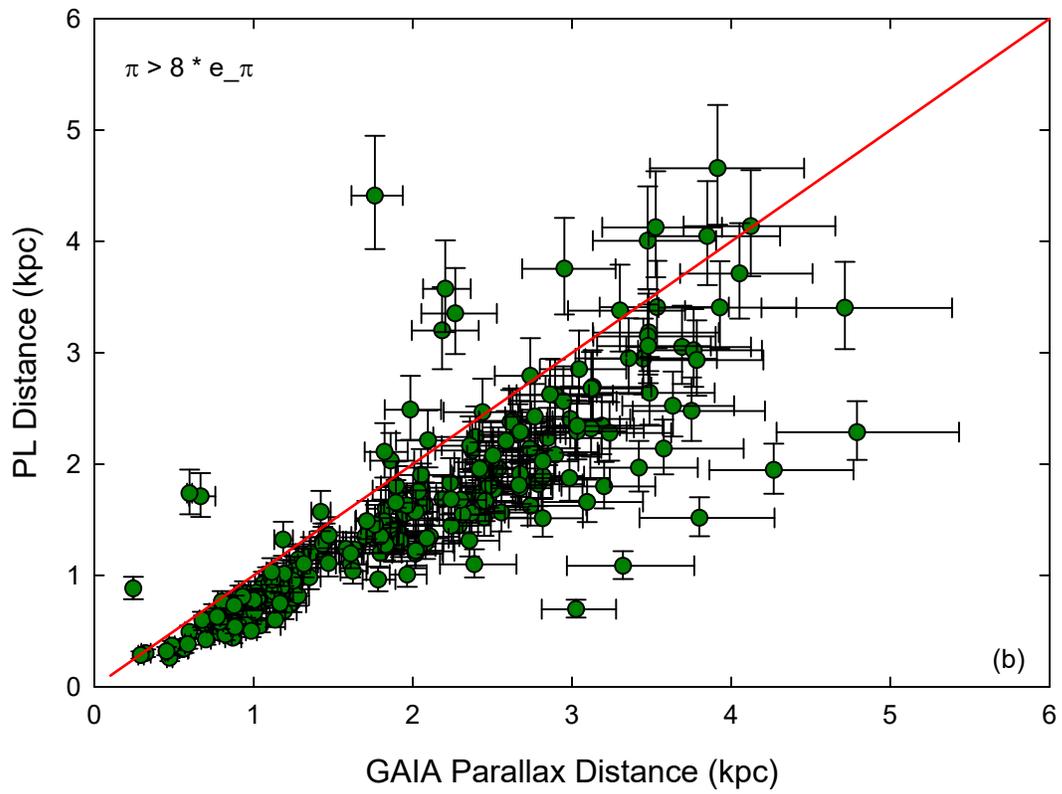

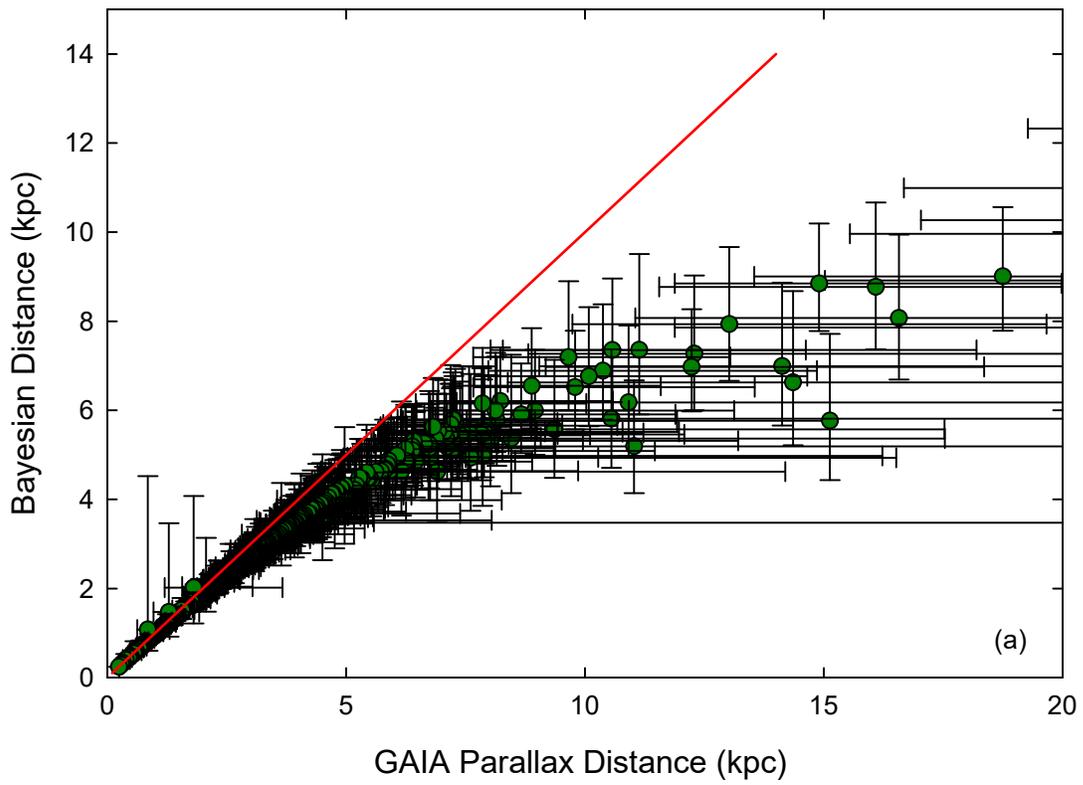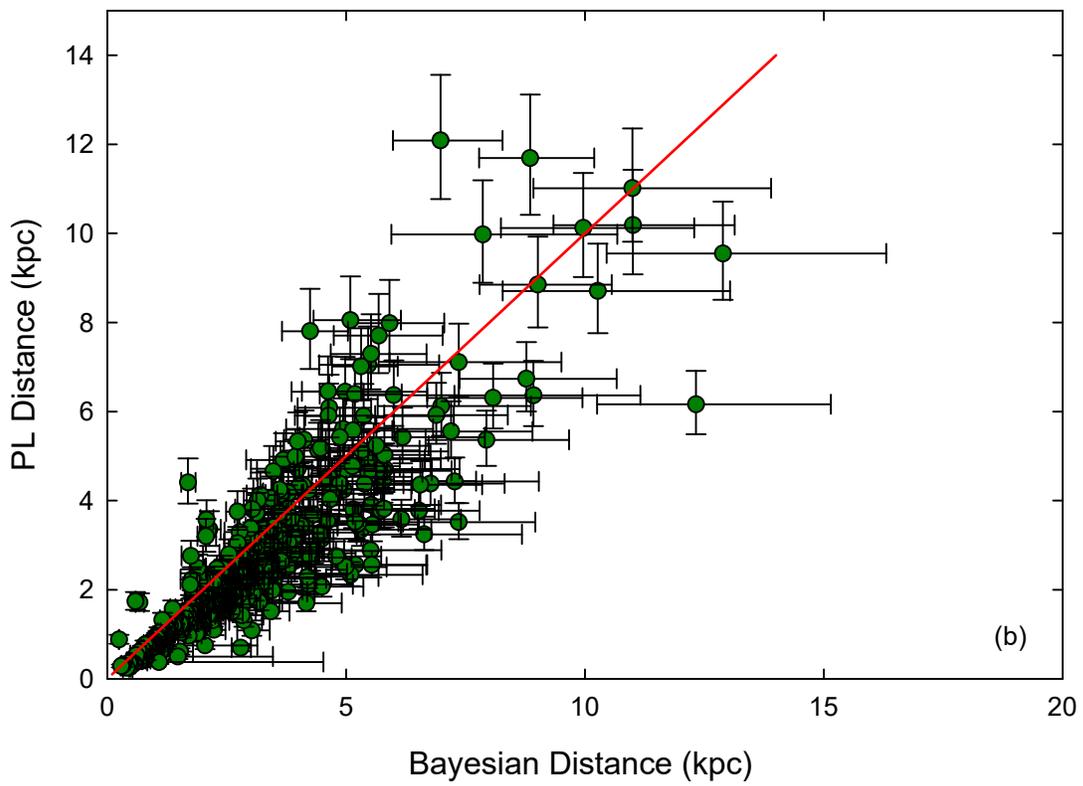

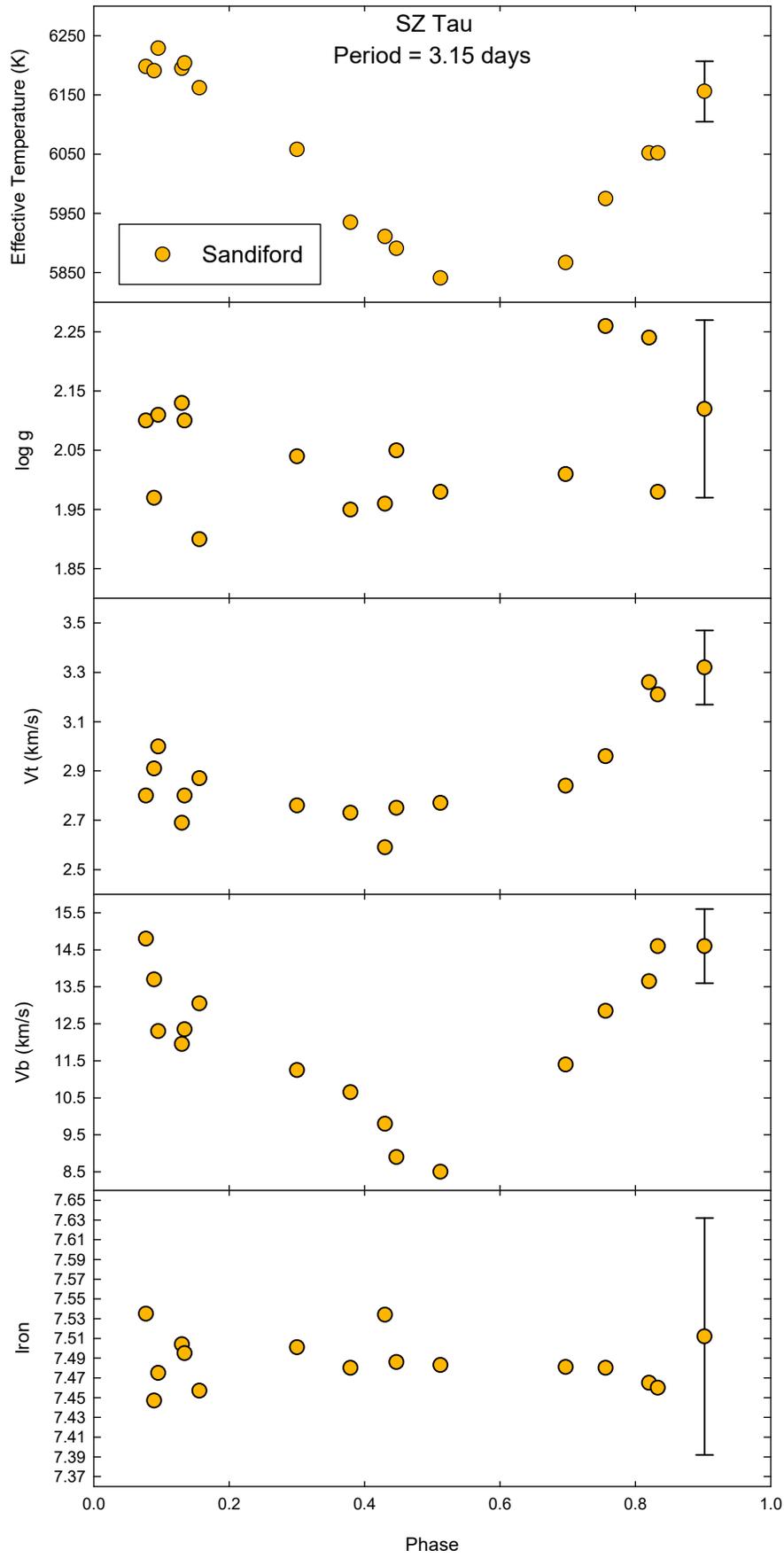

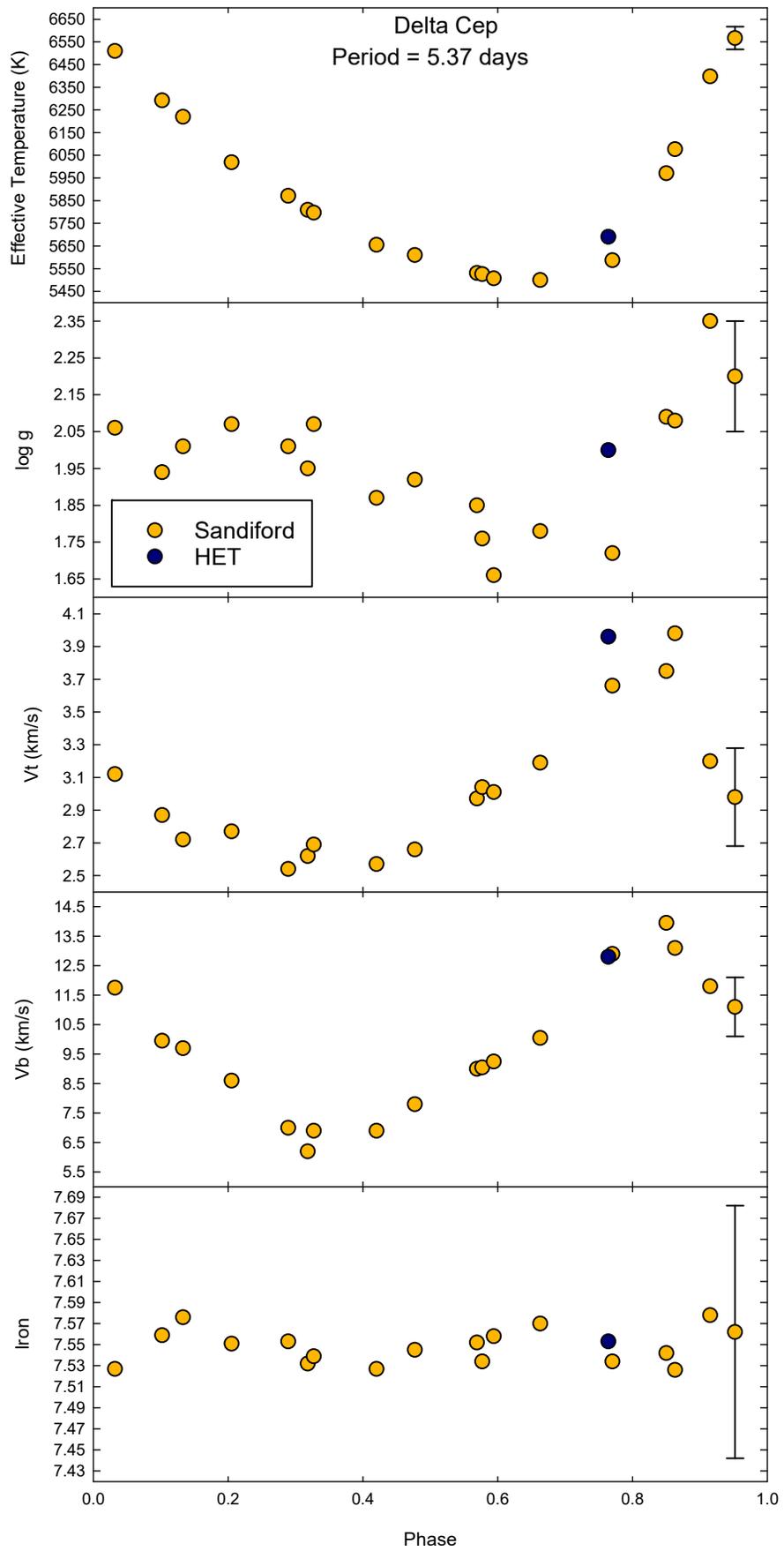

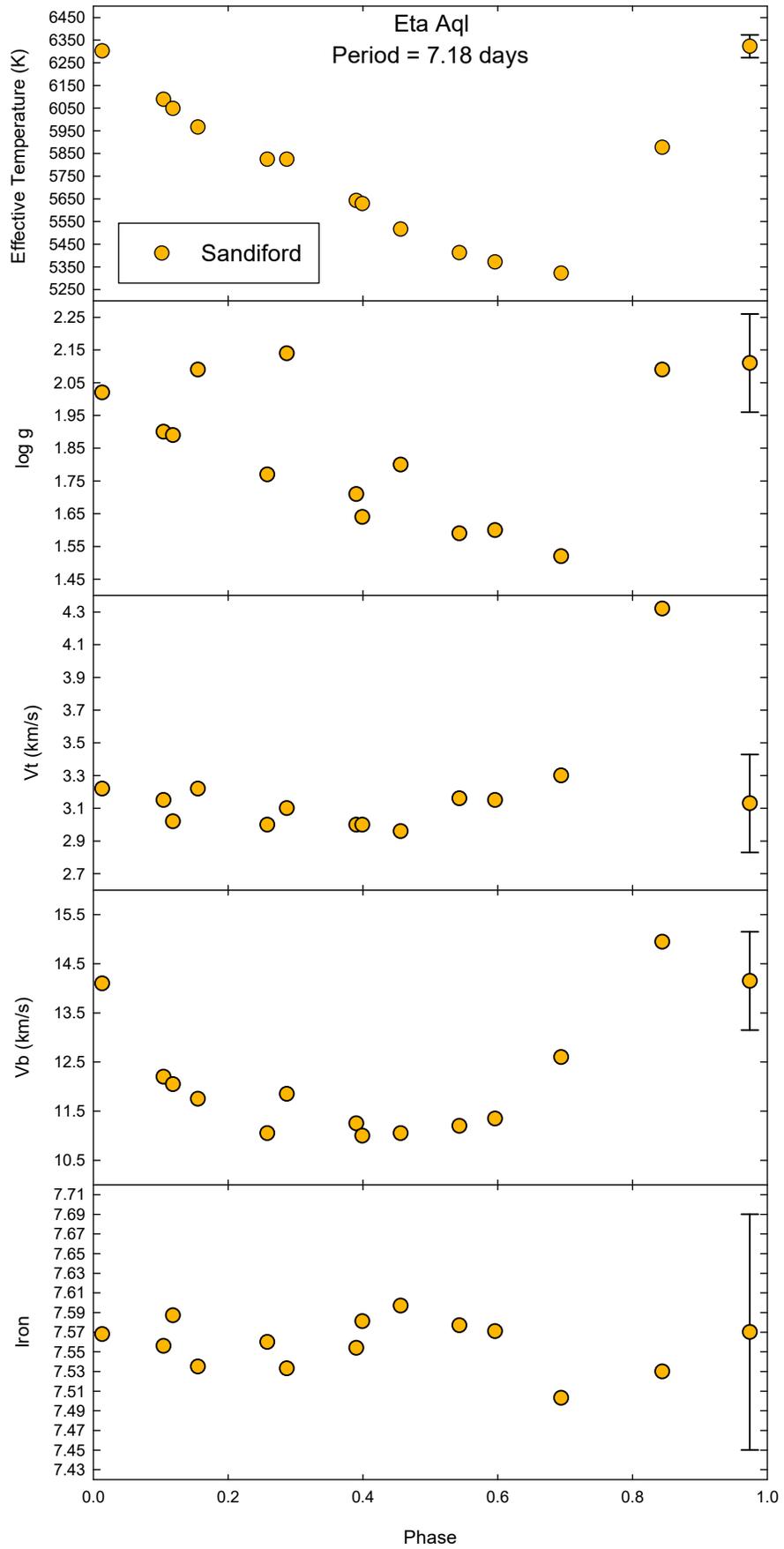

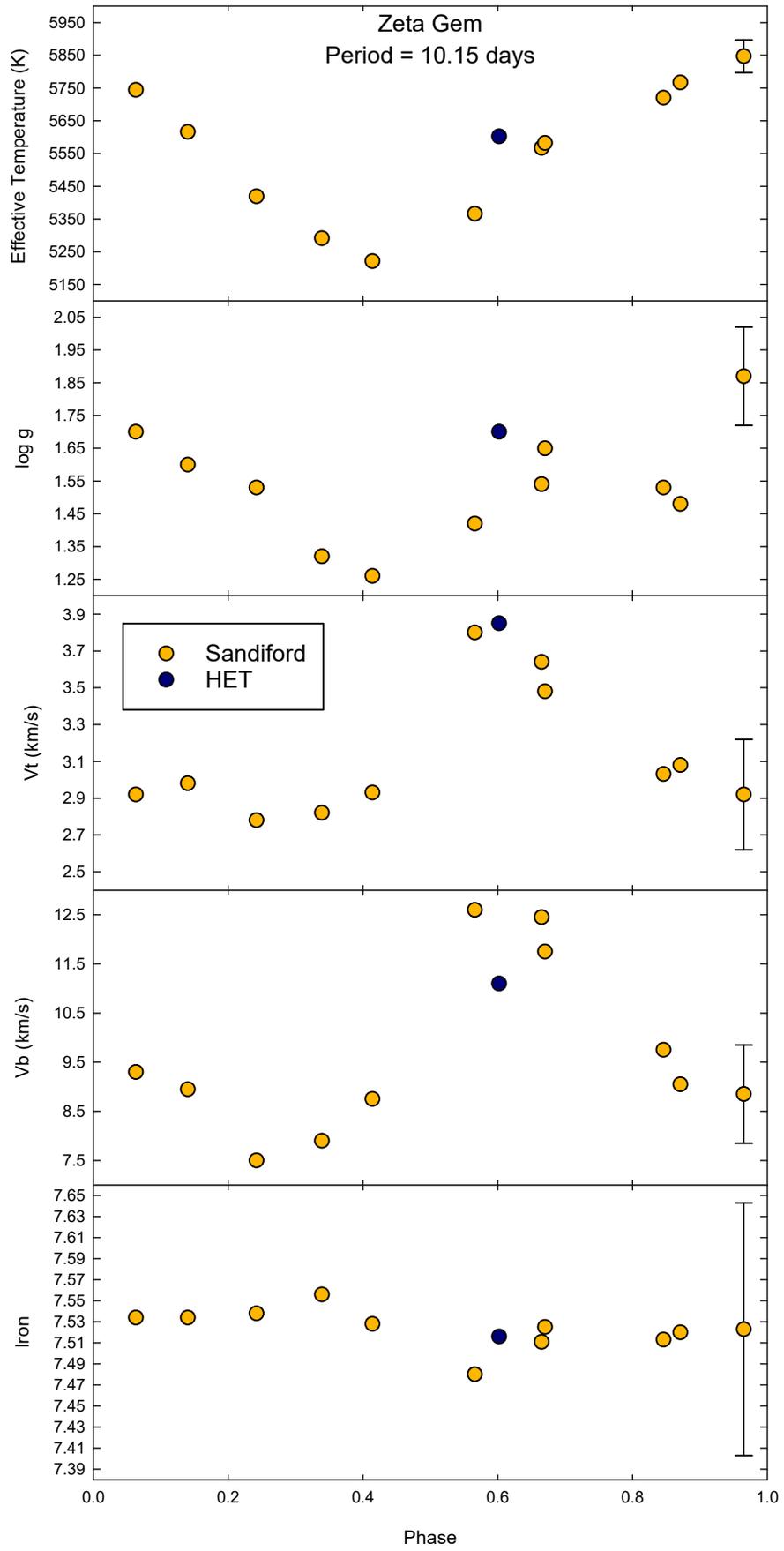

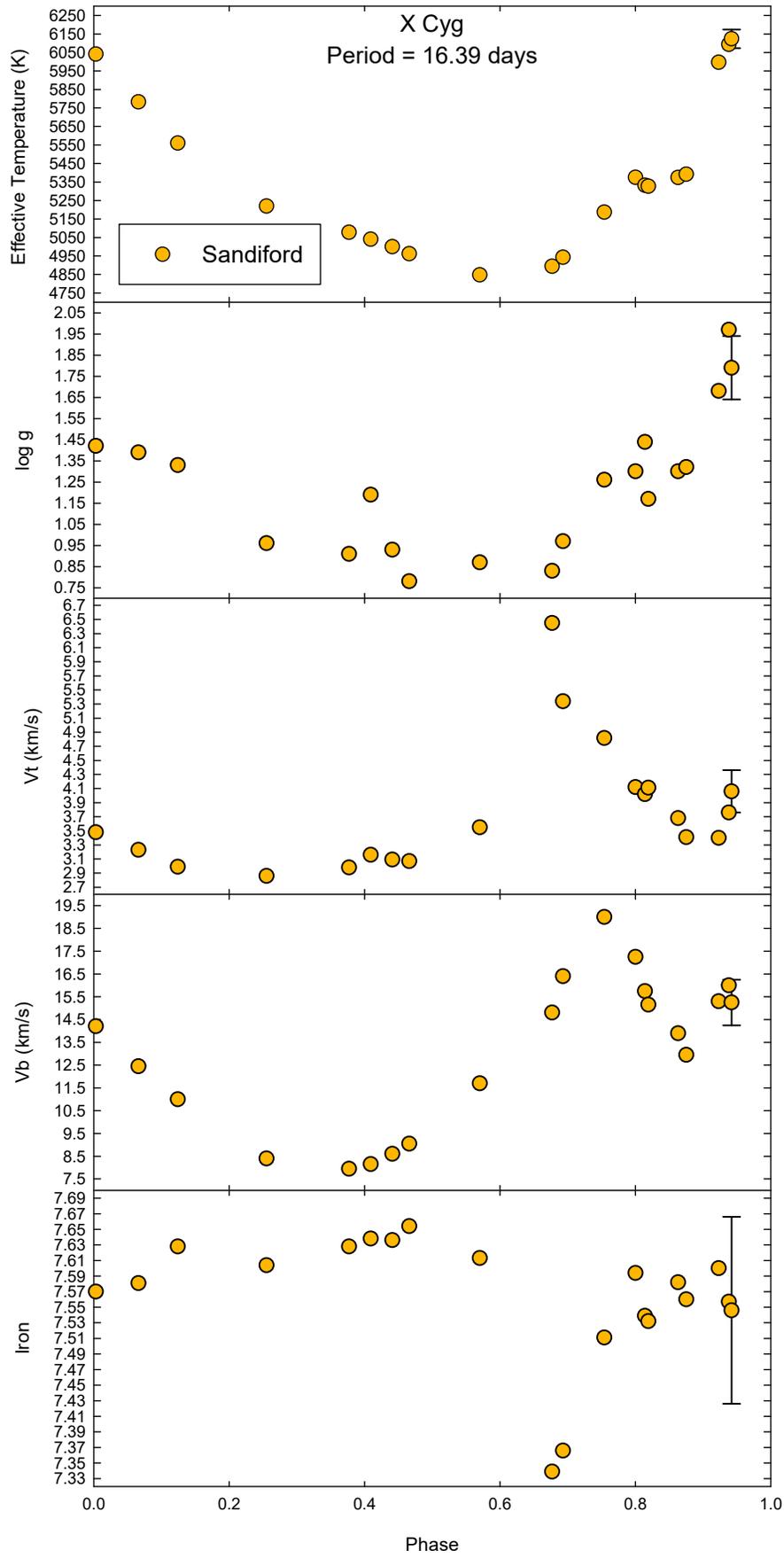

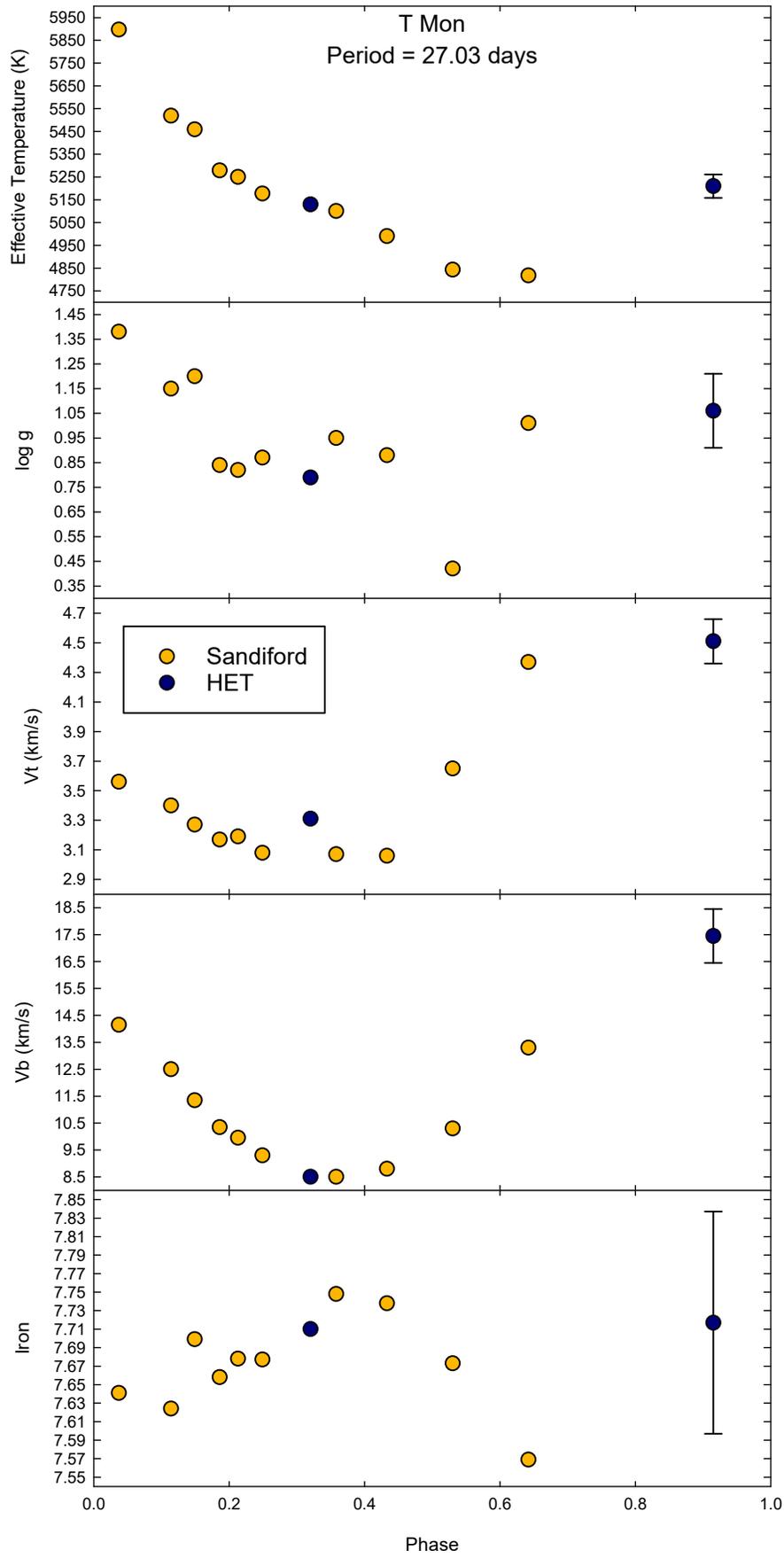

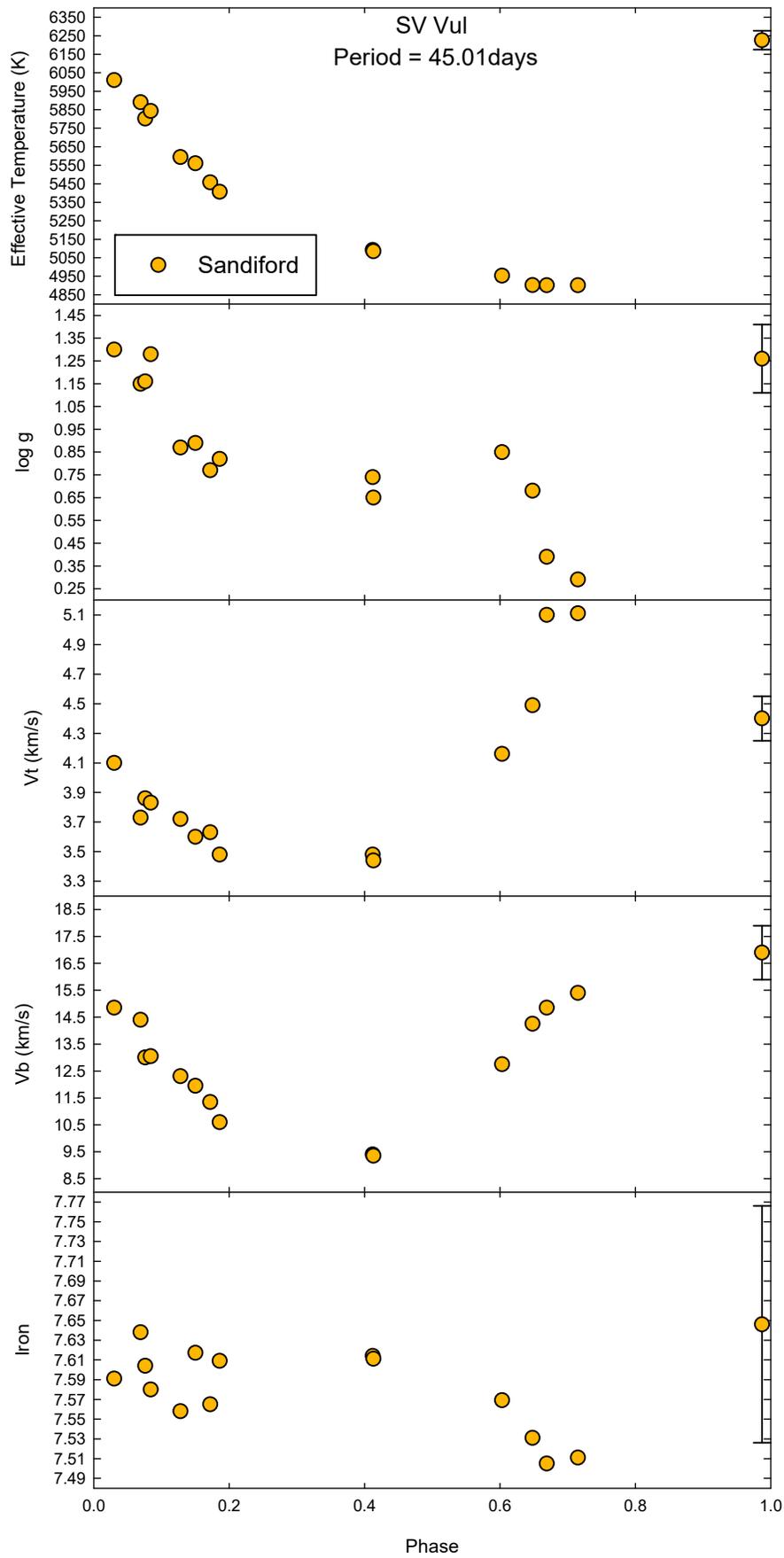

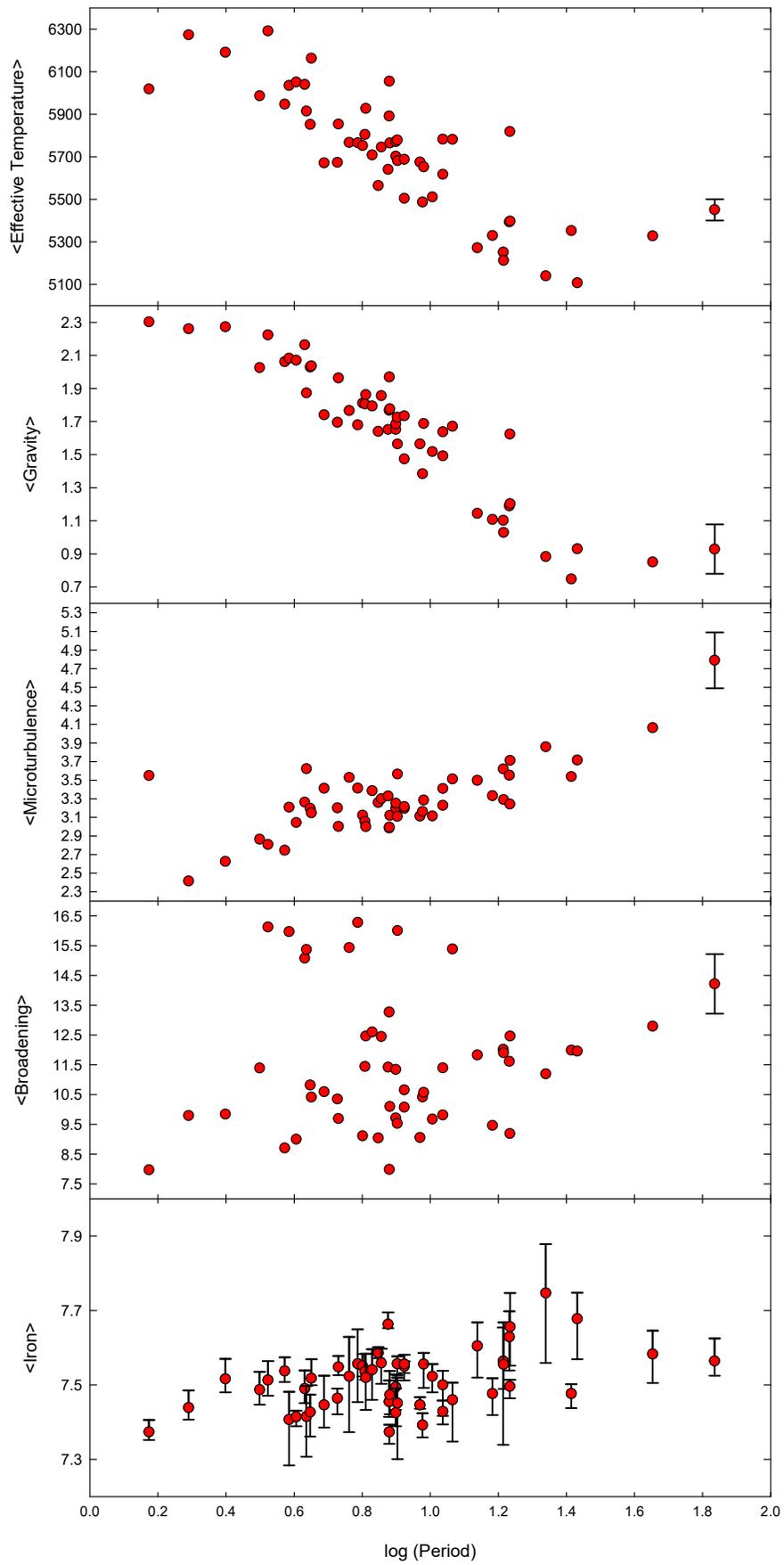

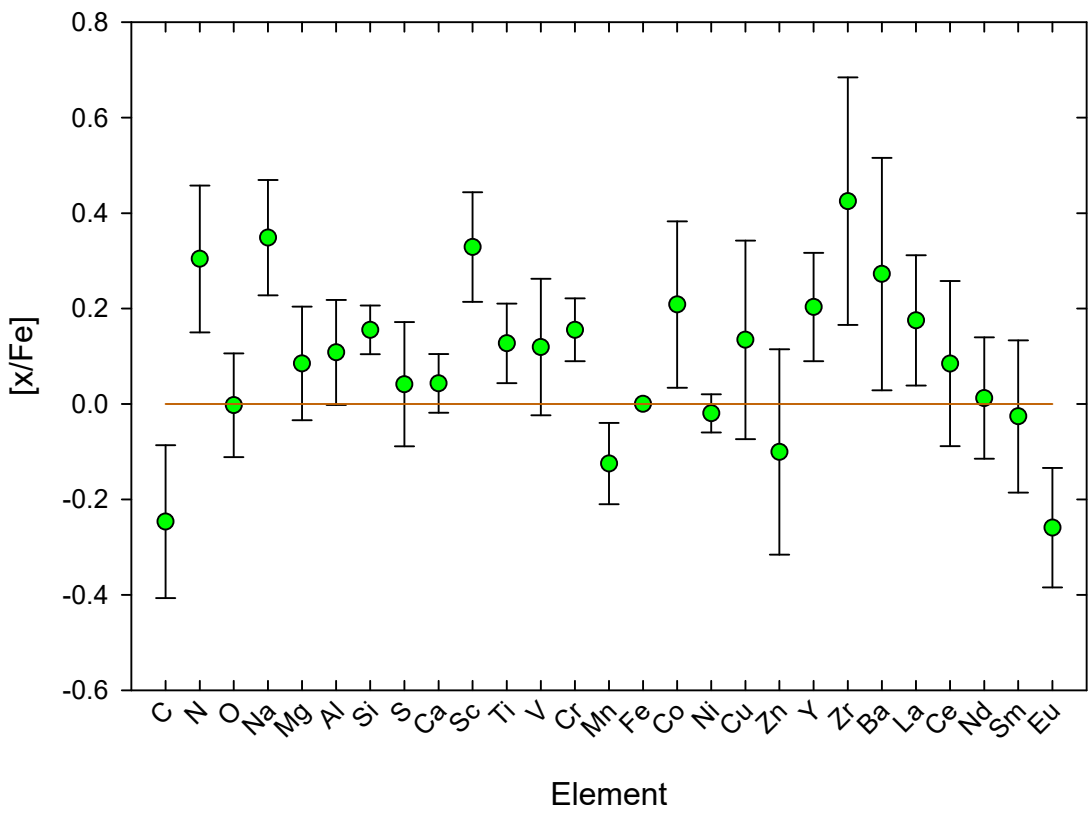

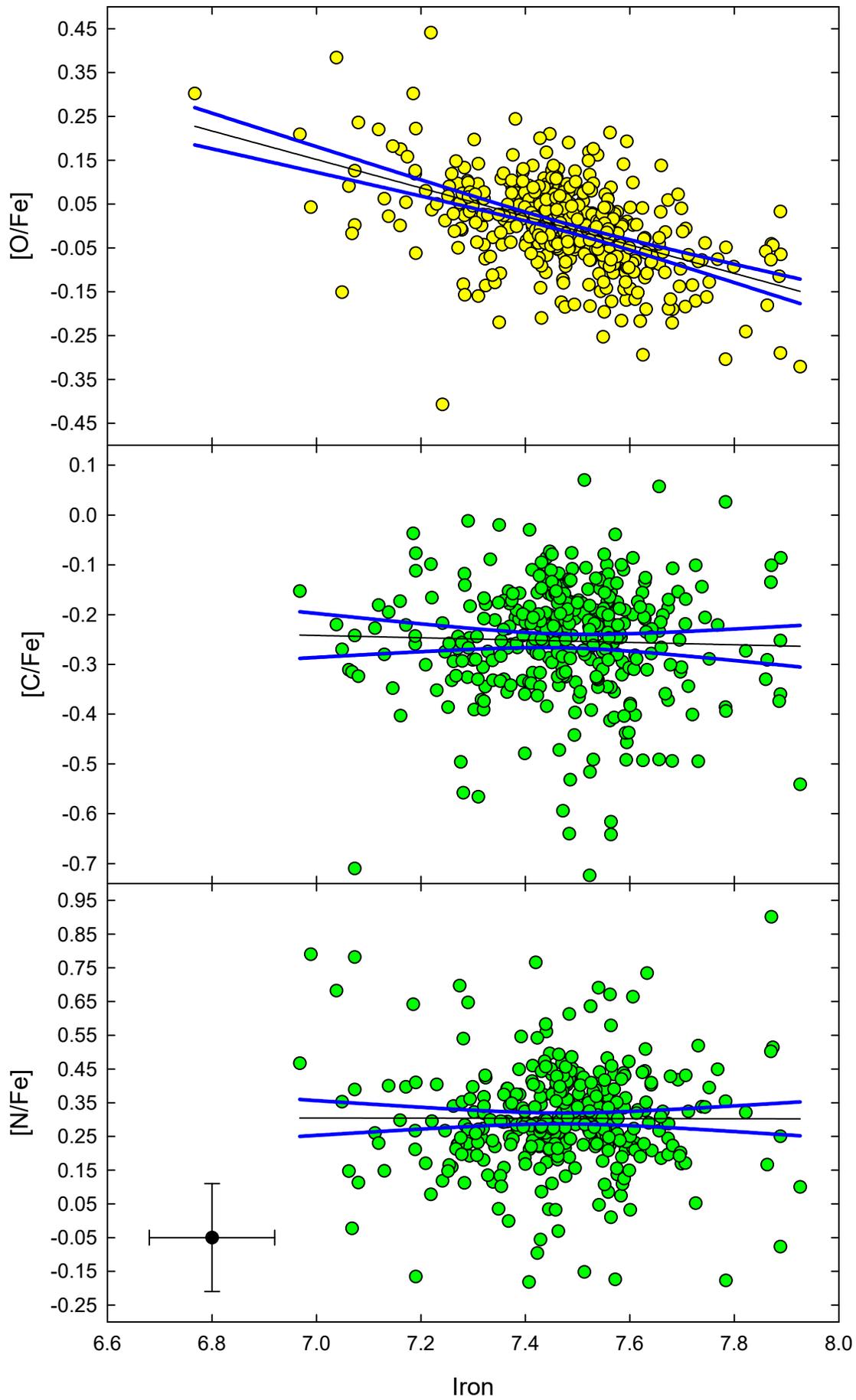

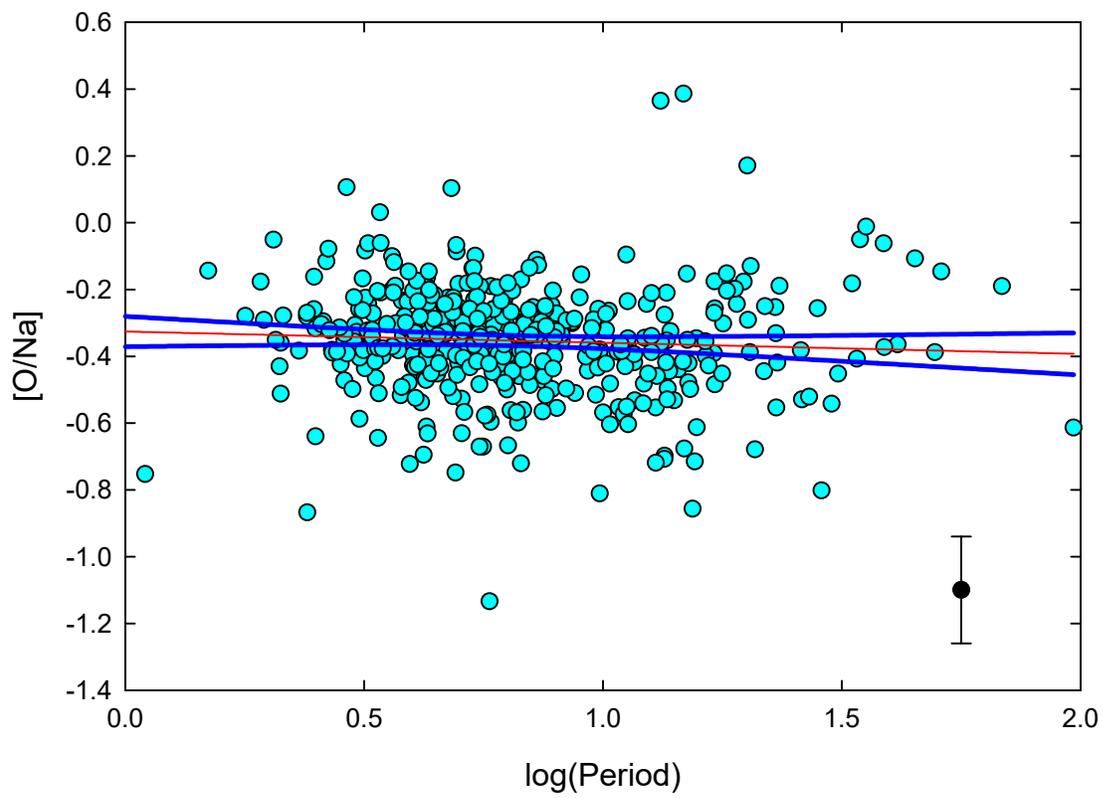

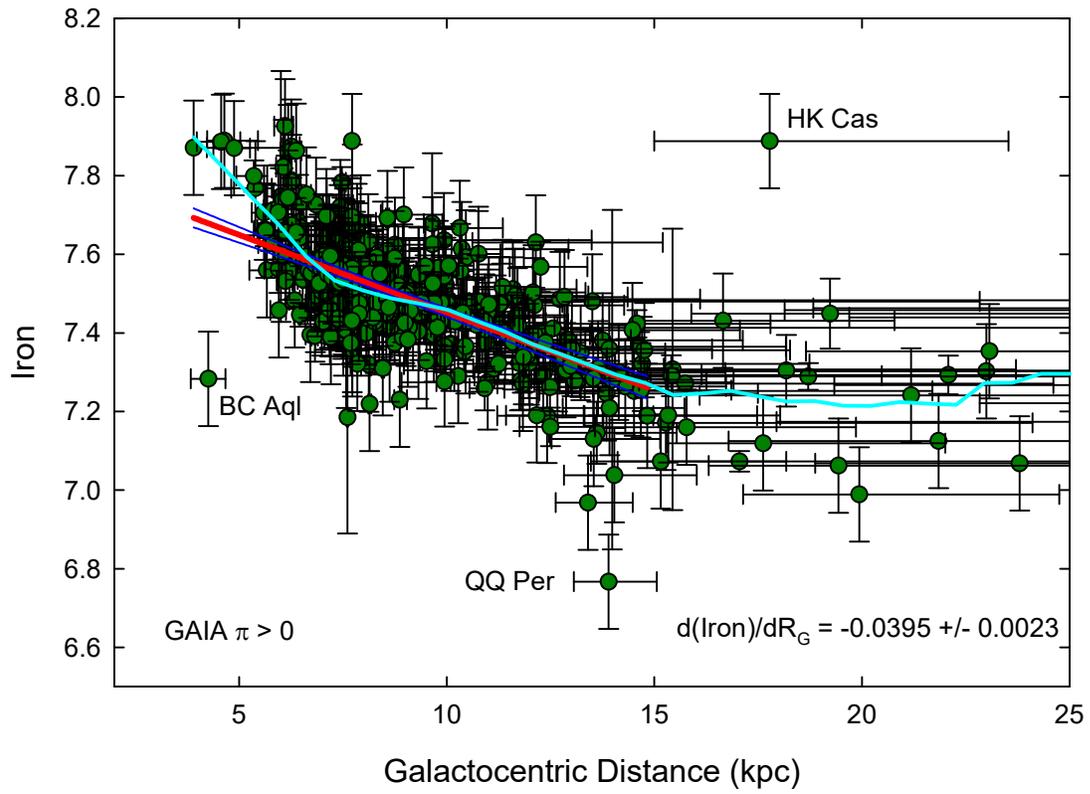
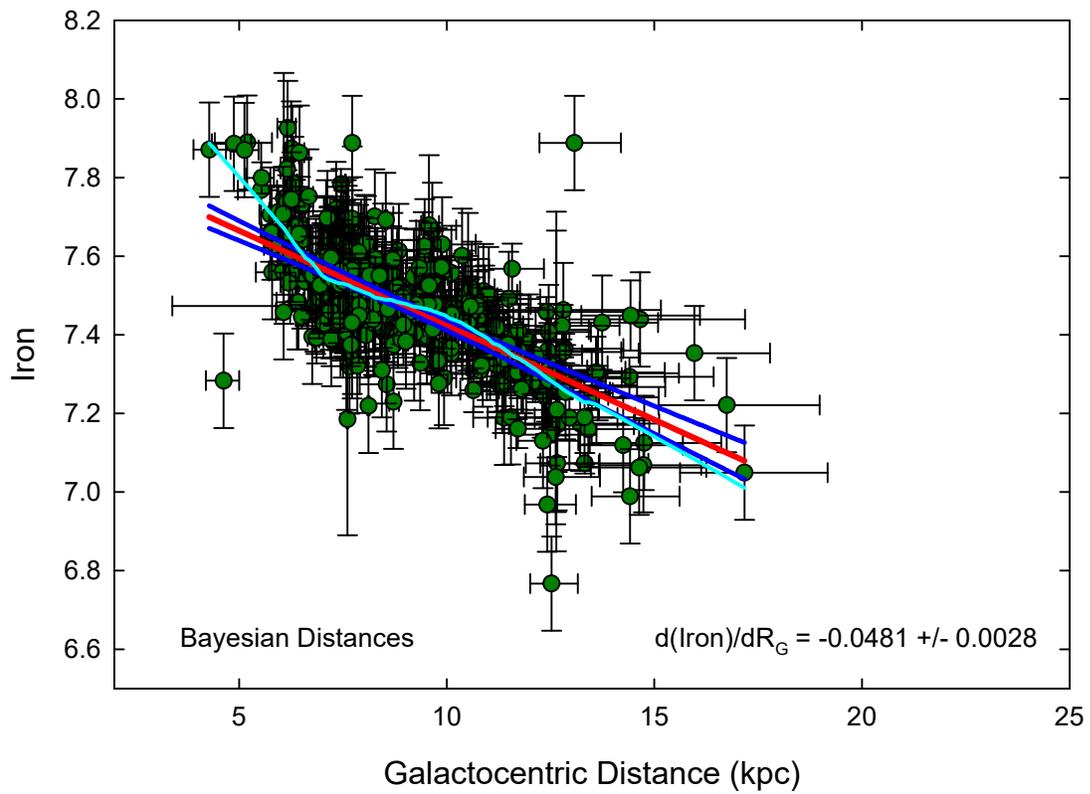

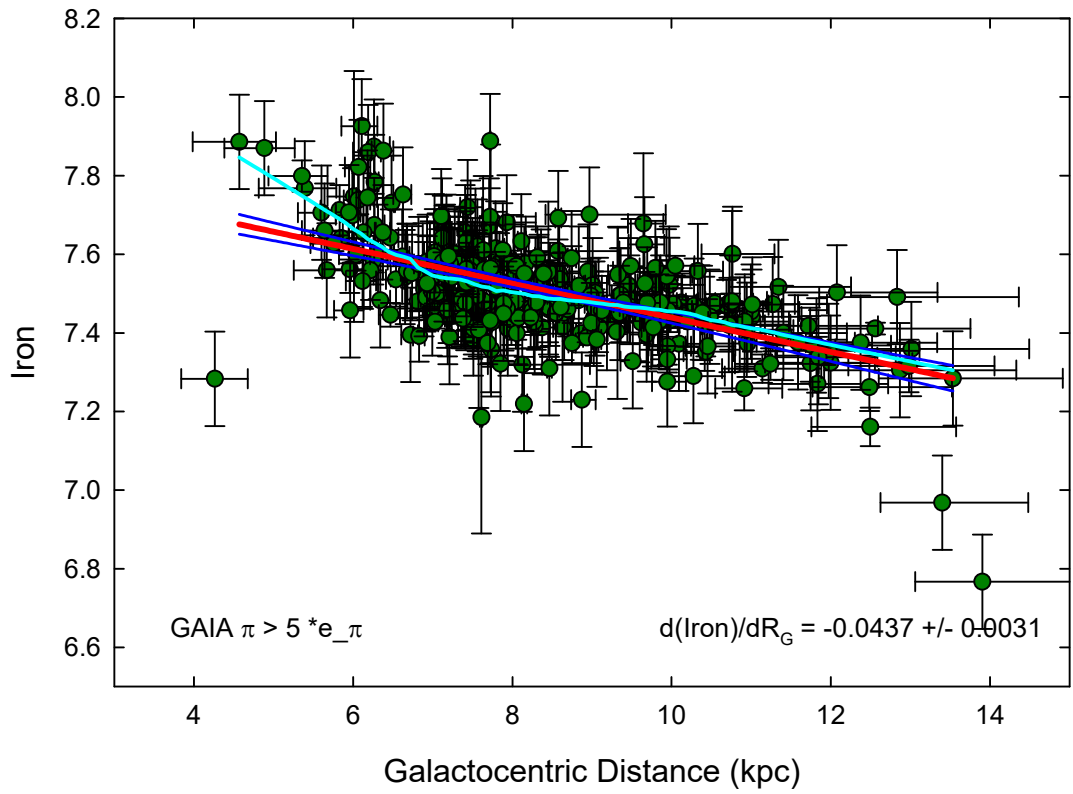

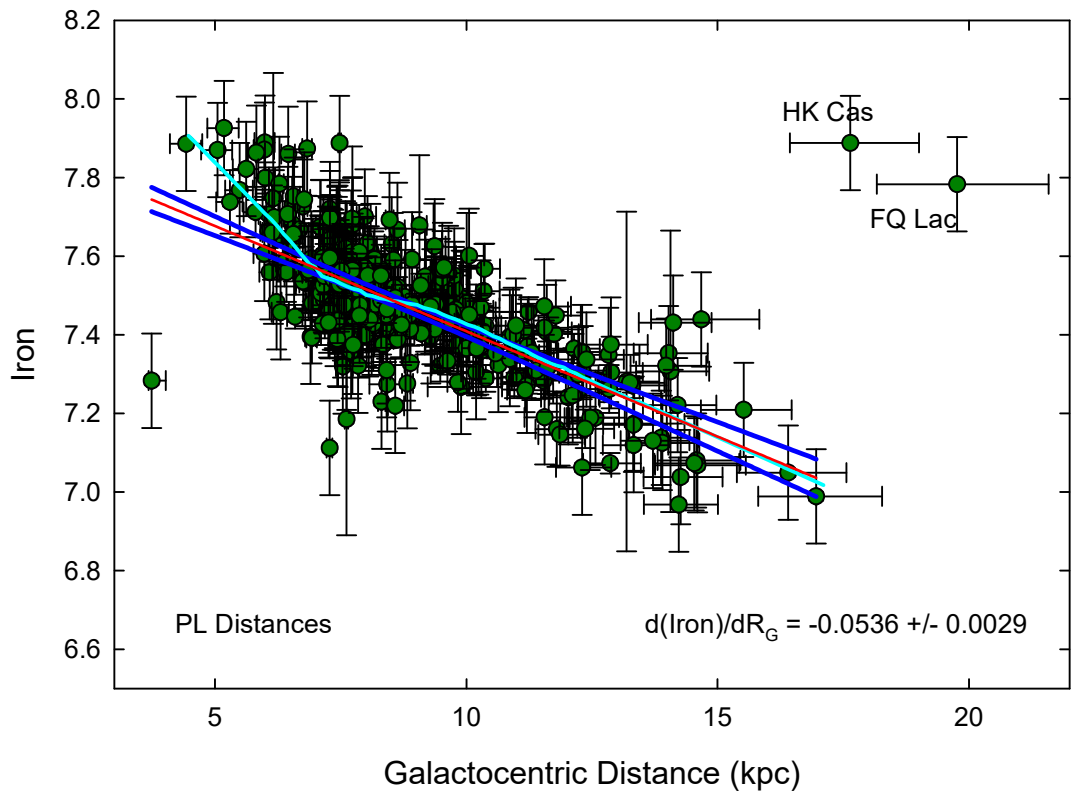

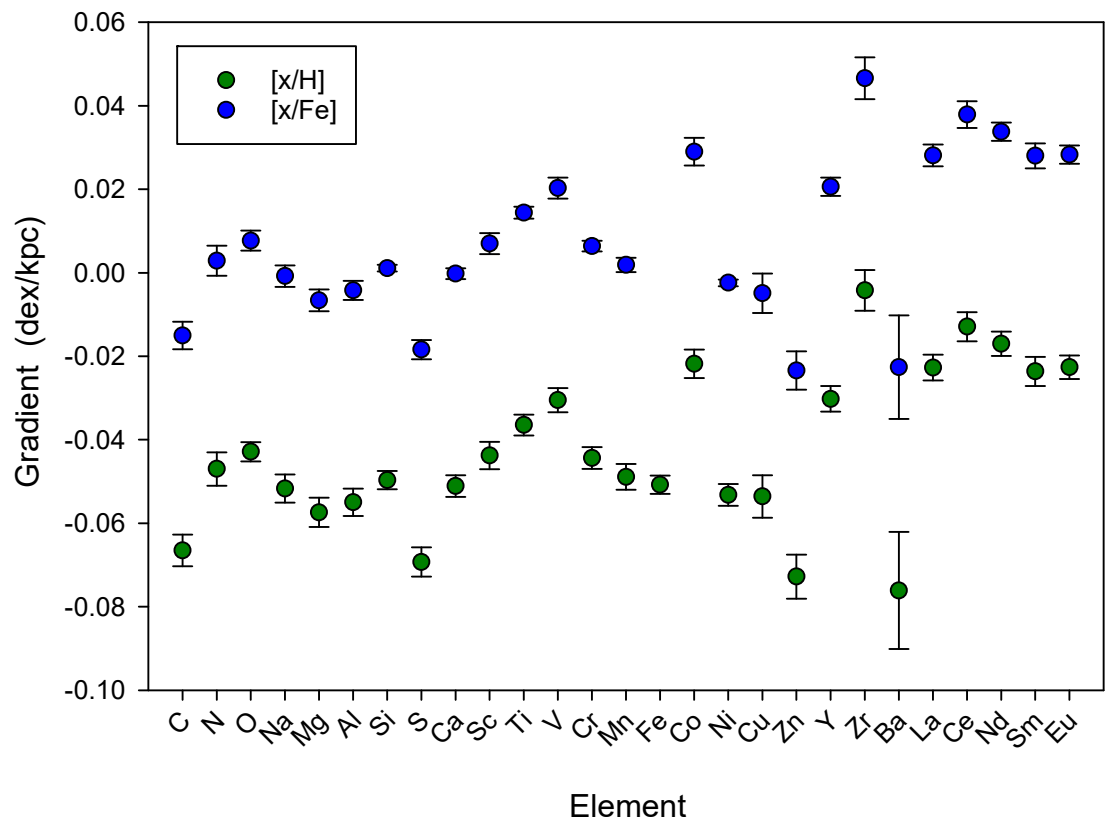